\documentclass[traditabstract]{aa}
\usepackage{graphicx}
\usepackage{txfonts}

 \newcommand{\mic}{$\mu$m}
 \newcommand{\mics}{$\mu$m~}

\def\tex {\ifmmode{{T}_{\rm ex}}\else{$T_{\rm ex}$}\fi}
\def\tmb {\ifmmode{{T}_{\rm mb}}\else{$T_{\rm mb}$}\fi}
\def\ci     {\ifmmode{{\rm C}{\rm \small I}}\else{C\ts {\scriptsize I}}\fi}
\def\hi     {\ifmmode{{\rm H}{\rm \small I}}\else{H\ts {\scriptsize I}}\fi}
\def\hh     {\ifmmode{{\rm H}_2}\else{H$_2$}\fi}

\def\ts     {\thinspace}
\def\kms    {\ifmmode{{\rm \ts km\ts s}^{-1}}\else{\ts km\ts s$^{-1}$}\fi}
\def\msol   {\ifmmode{{\rm M}_{\odot}}\else{M$_{\odot}$}\fi}
\def\lsol   {\ifmmode{{\rm L}_{\odot}}\else{L$_{\odot}$}\fi}
\def\zsol   {\ifmmode{{\rm Z}_{\odot}}\else{Z$_{\odot}$}\fi}
\def\etal   {{\rm et\ts al.}~}

\begin{document}

\title{Gas fraction and star formation efficiency at z $<$ 1.0
\thanks{Based on observations carried out with the IRAM 30m telescope.
IRAM is supported by INSU/CNRS (France), MPG (Germany) and IGN (Spain)}}

\author{F. Combes \inst{1}
\and
S. Garc\'{\i}a-Burillo \inst{2}
\and
J. Braine \inst{3}
\and
E. Schinnerer \inst{4}
\and
F. Walter \inst{4}
\and
L. Colina  \inst{5}
           }
\offprints{F. Combes}
\institute{Observatoire de Paris, LERMA (CNRS:UMR8112), 61 Av. de l'Observatoire, F-75014, Paris, France
\email{francoise.combes@obspm.fr}
 \and
Observatorio Astron\'omico Nacional (OAN)-Observatorio de Madrid,
Alfonso XII, 3, 28014-Madrid, Spain 
 \and
Laboratoire d'Astrophysique de Bordeaux, UMR 5804,Universit\'e Bordeaux~I, BP 89, 33270 Floirac, France
 \and
Max-Planck-Institut f\"ur Astronomie (MPIA), K\"onigstuhl 17, 69117 Heidelberg, Germany
 \and
Departamento de Astrofisica, Centro de Astrobiologia (CSIC/INTA), Torrej\'on de Ardoz, 28850 Madrid, Spain
              }

   \date{Received  September 2012/ Accepted February 2013}

   \titlerunning{CO in ULIRGs at 0.2 $<$ z $<$ 1.0}
   \authorrunning{F. Combes et al.}

   \abstract{After new observations of 39
galaxies at z $\sim$ 0.6-1.0 obtained at the IRAM 30m telescope,
we present our full CO line survey covering the redshift range 0.2 $<$ z $<$ 1.
Our aim is to determine the driving factors accounting for the 
steep decline in the star formation rate during this epoch. We study both the gas fraction,
defined as M$_{gas}$/(M$_{gas}$+M$_{star}$),
and  the star formation efficiency (SFE) defined by the ratio between
far-infrared luminosity and molecular gas mass (L$_{\rm FIR}$/M(\hh)), i.e.
a measure for the inverse of the gas depletion time.
 The sources are selected to be ultra-luminous infrared galaxies (ULIRGs),
with L$_{\rm FIR}$ greater than 10$^{12}$ \lsol\, and
experiencing starbursts. When we adopt a standard ULIRG CO-to-H$_2$ conversion
factor, their molecular gas depletion time is 
less than 100 Myr.  Our full survey has now filled the gap of CO
observations in the 0.2$<$z$<$1 range covering almost half of cosmic history.
The detection rate in the 0.6 $<$ z $<$ 1 interval is 38\%
(15 galaxies out of 39), compared to 60\% for the 0.2$<$z$<$0.6 interval. The average CO luminosity 
is L'$_{\rm CO}$ = 1.8$\times$10$^{10}$ K \kms\, pc$^2$, corresponding to an 
average \hh\, mass of  1.45$\times$10$^{10}$ \msol.
 From observation of 7 galaxies in both CO(2-1) and CO(4-3), a high gas excitation
has been derived; together with the dust mass estimation, this supports  the 
choice of our low ULIRG conversion factor between CO luminosity and \hh\,
for our sample sources. 
We find that both 
the gas fraction and the SFE significantly increase with redshift, by factors of $3\pm1$
from z=0 to 1, and therefore both quantities play an important role and complement each other
in cosmic star formation evolution. 
\keywords{Galaxies: high redshift --- Galaxies: ISM --- Galaxies: starburst ---
          Radio lines: Galaxies}
}
\maketitle

%---------------------------------------------------------------

\section{Introduction}

The star formation history (SFH) of the Universe shows a steep decline by a factor 10
between z=1 and 0, after a peak of activity around z=1-1.5,
well within the first half of the Universe's age  
(Madau et al 1998, Hopkins \& Beacom 2006).
During the second half of the Universe, fundamental changes occur not only in the star formation rate,
but also in both the galaxy interaction/merger rate (Le F\`evre \etal 2000, Conselice \etal 2009, Kartaltepe \etal 2010)
and the galaxy morphology (Sheth \etal 2008, Conselice \etal 2011), as the well known local Hubble sequence
 is just now assembling.

The main physical processes causing the decline in the SFH are not well understood. 
One frequently invoked factor is the gas content, or gas fraction at a given stellar mass, since
gas is the fuel for star formation.  Locally, the gas fraction for giant spirals is 
about 7-10\% (Leroy \etal 2008, Saintonge \etal 2011a), while at z$\sim$1.2 it increases
to 34$\pm$5\% and at z$\sim$2.3 to 44$\pm6$\% (Tacconi \etal 2010, Daddi \etal 2010). 
However, the average surface and volume gas densities should be  more relevant
factors, and there is evidence of a wide range in 
gas properties of galaxies at each redshift: 
molecular gas extents, mean gas densities and star
formation efficiencies (Daddi \etal 2008). The average molecular gas density can be traced by the CO excitation,
measured from line ratios of the CO ladder (e.g. Weiss \etal 2007).
Another intervening factor is an external dynamical trigger of star formation,
such as galaxy interactions or accretion by cold gas. Both effects are expected to increase with redshift,
up to z=2-3.

Galaxy mergers provide the violent gravity torques able to drive the gas quickly
to galaxy centers and to trigger starbursts. Locally
 Ultra-Luminous Infra-Red Galaxies (ULIRGs), with 
L$_{\rm FIR}  > $  10$^{12}$ L$_\odot$, are in the majority starbursts caused
by galaxy major mergers (e.g. Sanders \& Mirabel 1996,  Veilleux \etal 2009). 
The overall gas fraction in those systems is not high, and they are expected
to sustain such a high rate of star formation for only a short duration, on the order
of 100Myr. The depletion time scale of massive star-forming galaxies, with a high gas fraction but
low star formation efficiency (SFE), is on the other hand on the order of one Gyr or more (Leroy \etal 2008,
Bigiel \etal 2008).
These galaxies are dubbed as falling on the ``main sequence'' (as defined below) of star-forming galaxies 
(Noeske \etal 2007), or in a  ``normal'' star formation phase.

These two modes of star formation: main sequence (MS) and merger-induced starbursts
(SB) have been been discussed as a function of stellar mass. 
At both high- and low-z, several studies (Noeske 2009; 
Rodighiero \etal 2011; Wuyts \etal 2011 ) have shown a broad correlation between
the star formation rate (SFR) and the stellar mass (M$_*$). The bulk of star-forming galaxies follow  the ``normal''
mode of star formation in the center of correlation, and
define the SFR-M$_*$ as ``main sequence''. The galaxies in the upper envelope of this main sequence
have higher SFR at a given mass. These starbursts are rather rare, and 
represent only 10\% of the cosmic SFR density at z $\sim$ 2  (Rodighiero \etal 2011).
  They could be either a common phase of ``normal'' galaxy life, or the end phase leading
to a massive early-type galaxy, where star formation is quenched. 

The specific SFR (i.e. star formation per unit mass) sSFR 
is found either to decrease slightly with  $M_*$ ($\propto M_*^{-0.2}$,
Noeske \etal 2007), or to be a constant with  $M_*$ (Wuyts \etal 2011).
The global SFR decreases with cosmic time as SFR $\propto$ (1+z)$^{2.7}$ out
to z=1-2. At each epoch, there is a population of
non-star-forming (quiescent) galaxies of high mass and high Sersic index.
These galaxies might correspond to quenched galaxies and their average stellar mass increases with z.
The starbursts at the top of the main sequence have a morphology
intermediate between the MS and quiecent objects (Wuyts \etal 2011).

To better clarify the role of each physical process in the SFH decline in the
second half of the Universe's history, it is of prime importance to determine
the gas content, spatial extent and surface density of star-forming objects 
between redshifts 0.2 and 1.  To do so we have begun a CO line survey
of ULIRGs in this still unexplored redshift range 
(e.g. Combes \etal 2011, hereafter Paper II, where we focused on objects in the
redshift range 0.2$<$z$<$0.6), and we now complement it
with  0.6 $<$ z $<$ 1 objects.

Very little is known about the molecular gas content of
galaxies in this redshift range because of observational difficulties.
Paradoxically, it is often easier to study higher redshift galaxies,
because the global CO line flux increases almost like the square 
of frequency for high-J lines (Combes \etal 1999).
 In the most favorable 3mm atmospheric window,
between 81 and 115\,GHz, all redshifts can be observed with 
at least one line of the CO rotational ladder, except between z=0.4 and 1.
 For galaxies in this range, CO lines must be searched for in the 2mm
or 1mm range, and several CO lines are out of reach.

In this paper we focus on a sample of 39 
IR-luminous galaxies in the range 0.6 $<$ z $<$ 1 and detected either
by IRAS or Spitzer. From their
stellar mass and derived sSFR, these objects are mostly in the starburst phase
on the upper envelope of the main sequence.
Their sSFR  helps us to determine which CO-to-H$_2$ conversion ratio
 to use, and therefore provides the best estimates of the true star formation efficiency.
There appears to be a variation in conversion ratios by a factor $\sim$ 6
between the two extreme classes of objects. In local ULIRGs where
a compact nuclear starburst is typically mapped, the CO emission is 
much stronger for a given H$_2$ mass, and the generally adopted
 ratio between M(\hh) and L'$_{\rm CO}$, expressed in units of M$_\odot$ (K \kms\, pc$^2$)$^{-1}$,
is $\alpha$=0.8 instead of the adopted Milky Way ratio
of $\alpha$=4.6 (Solomon \etal 1997).
However, it is still possible that this ratio underestimates the molecular
mass in ULIRG (e.g. Papadopoulos \etal 2012).

 To determine the extent of the molecular gas, interferometric maps 
are required, and we are also mapping some of the detected galaxies 
with the IRAM Plateau de Bure interferometer (PdBI) 
(Combes \etal  2006, Paper I). This first map of a galaxy at z=0.223 revealed
that at least half of the emission is extended on scales of 25-30kpc.
 We will present other PdBI maps in a future article, so the focus of the present
paper is the second part of our survey.
The sample is described in Sect. \ref{sample} and the observations
in  Sect. \ref{obs}. Results are presented in  Sect. \ref{res} and
discussed in Sect. \ref{disc}. 

\begin{figure}[h]
\label{fig:sample}
\includegraphics[angle=-90,width=8cm]{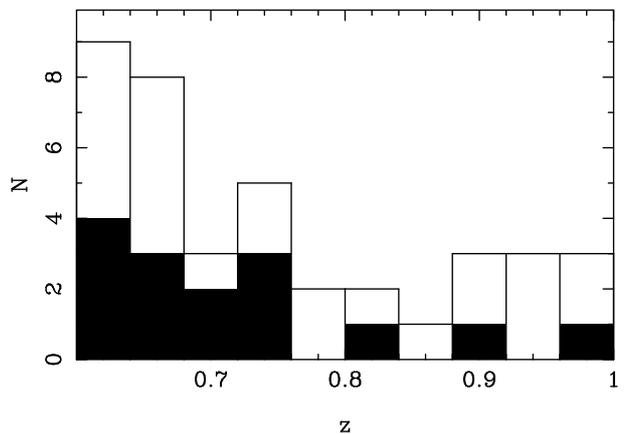}
\caption{Redshift histogram of the presented sample (see Table \ref{tab:sample}) 
at redshifts between 0.6 $<$ z $<$ 1.0.
The sources have been selected based on their 60 \mic\, flux from IRAS, or 70 \mic\,  flux from
Spitzer-MIPS. The open histogram indicates CO-undetected sources,
while the filled histogram shows detections.}
\end{figure}

\section{The sample}
\label{sample}

We used the same criteria as in Paper II (except for the redshift range),
and selected ULIRG galaxies
with log $L_{\rm FIR} /\lsol >$ 12.45,
between 0.6 $<$ z $<$ 1, with a declination greater than -12$^\circ$,
 a spectroscopic redshift, and detected at 
 60\mic\, (IRAS) or at 70\mic\, (Spitzer).
 Our initial sample during the first run contained 39 objects; however, 
observations of some galaxies could not be used
because of baseline oscillations caused by their
strong continuum flux, or because of noisy spectral parts of the receivers.
 We have therefore removed those galaxies from our sample and replaced them
with slightly lower far-infrared luminosities, but keeping the
ULIRG (L$_{\rm FIR}>10^{12}\lsol$) criterion. Among the 39 sources, 32 have 
SDSS images; they often appear to be point sources and, 
in a few cases, show perturbed and interacting systems.
Ten sources have been imaged in the near-infrared with sub-arcsec seeing
by Stanford \etal (2000).  HST images are available for about half
of the sample (17 objects). From the available imaging, it is possible
to identify morphological perturbations or minor companions
in about 20\% of the cases, and obvious strong interactions in about 30\%.
The remaining half of the objects appear unperturbed. This high 
frequency of unperturbed morphologies, in comparison
with our sample between 0.2 $<$ z $<$ 0.6, is at least in part due to the 
lack of high spatial resolution and high sensitivity optical imaging for the higher-z sample. 

The name of the sources, their coordinates and redshifts are given in
Table \ref{tab:sample}.
Out of the 39 objects in the sample, 15
were detected, corresponding to a detection rate of 38\%.
Figure  \ref{fig:sample} displays their distribution in redshift.

\begin{figure}[!t]
\resizebox{8cm}{!}{\includegraphics[angle=0,width=8cm]{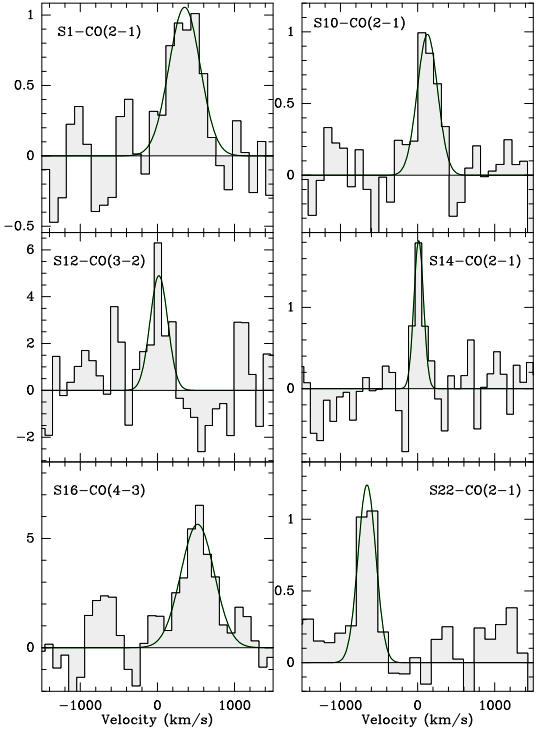}}
\caption{CO spectra of the detected galaxies. Zero velocity 
corresponds to the optically determined redshift,
listed in Table  \ref{tab:sample}. According to the redshift, the lowest
frequency observable is either the CO(2-1) or CO(3-2) line.
One source (S16) is detected in CO(4--3) but not in CO(2-1),
as indicated in Table \ref{tab:lines}.
The vertical scale is T$_{mb}$ in mK.}
\label{fig:spec1}
\end{figure}

\begin{figure}[!t]
\resizebox{8cm}{!}{\includegraphics[angle=0,width=8cm]{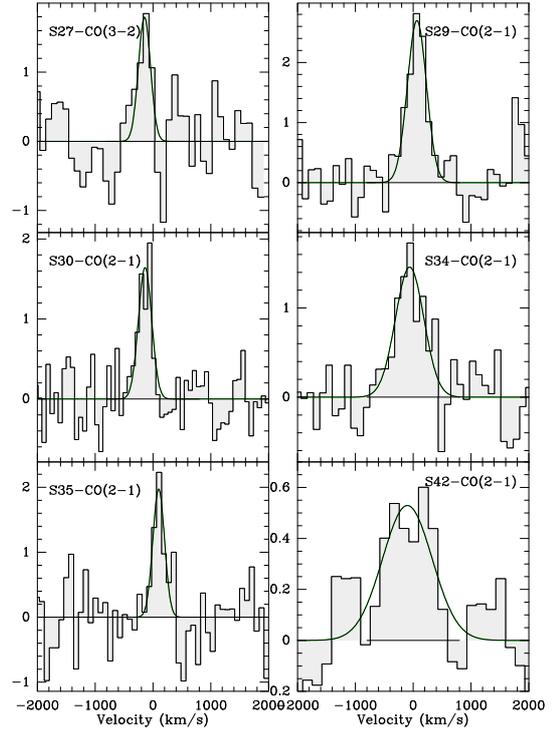}}
\caption{ Same as Fig. \ref{fig:spec1} for additional galaxies.}
\label{fig:spec2}
\end{figure}

\begin{figure}[!t]
\resizebox{8cm}{!}{\includegraphics[angle=0,width=8cm]{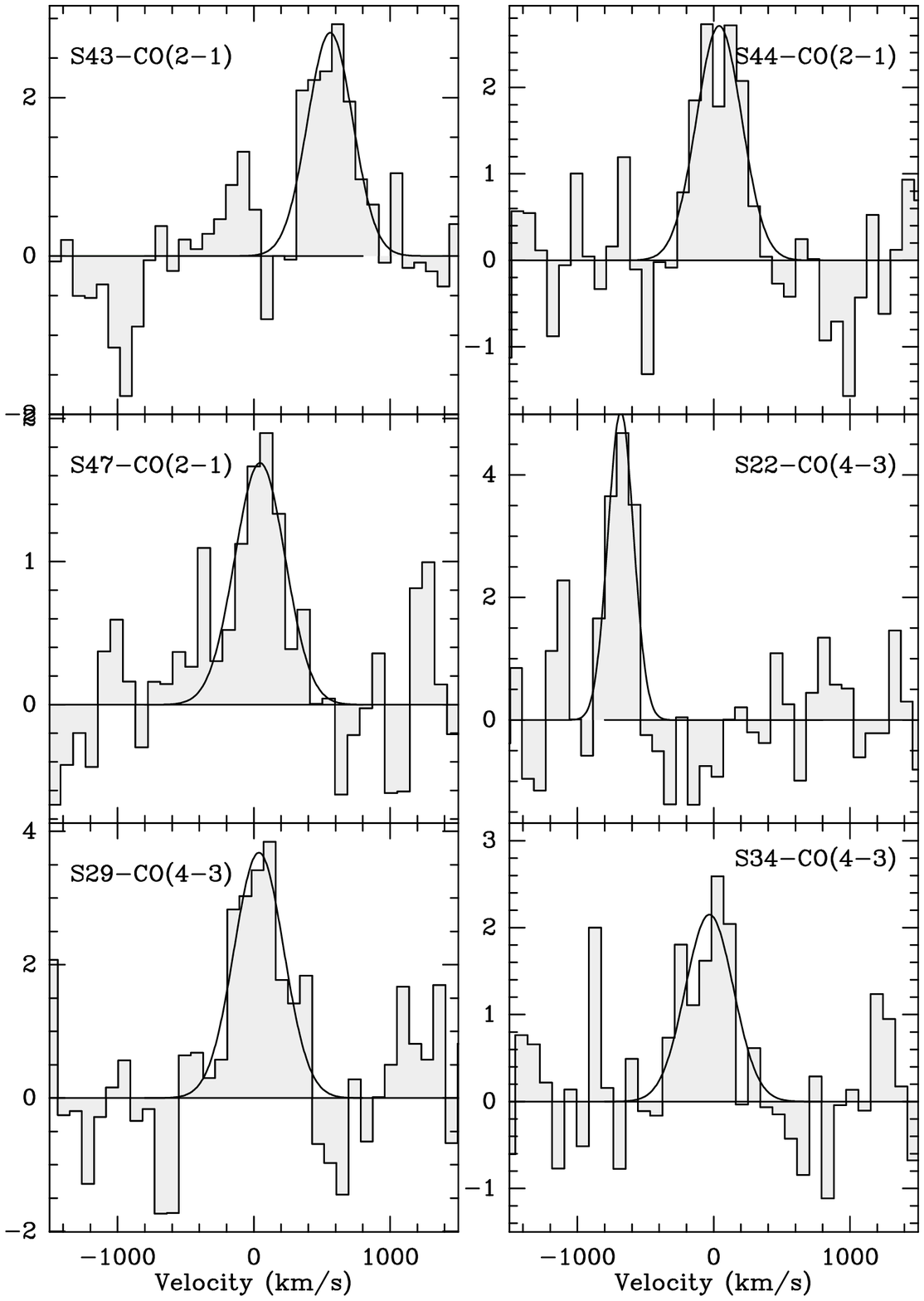}}
\caption{Same as Fig. \ref{fig:spec1}  for the remaining 3 galaxies. The final 3 plots
are CO(4-3) spectra from some sources already shown in CO(2-1).}
\label{fig:spec3}
\end{figure}

The majority of the sources have IRAS fluxes at 60 and 100\mic, and
their far-infrared fluxes F$_{\rm FIR}$ are 
 computed as 1.26 x 10$^{-14}$ (2.58 S$_{60}$+S$_{100}$) W m$^{-2}$ 
(Sanders \& Mirabel 1996). The far-infrared luminosity between 40 and 500\mics 
is then L$_{\rm FIR}$  = 4 $\pi$ D$_L^2$ CC F$_{\rm FIR}$, where 
D$_L$ is the luminosity distance, and
CC the color correction, CC=1.42. Some of the sources instead have
Spitzer-MIPS fluxes at 70 and 160\mic. We chose to compute their
F$_{\rm FIR}$ fluxes similarly, with a linear combination of these two
fluxes, since all sources are ultraluminous objects; i.e., their SEDs fall into
the same category, and this gives comparable results 
to those of SED fitting (e.g. Symeonidis \etal 2008).
 The FIR-to-radio ratio 
$q$=log([F$_{\rm FIR}$/(3.75 10$^{12}$ Hz)]/[f$_\nu$(1.4 GHz)])
has been computed for sources where radio data are available;
 the radio fluxes are listed in Table \ref{tab:lines}.
If excluding the radio galaxies 3C280 and 3C289 (S18 and S20),
the average is $q$=1.7, close to the nominal value for ULIRGs (Sanders \& Mirabel 1996). 
 The SFRs are all 
above 200 $\rm M_{\odot}yr^{-1}$, and the sample average 
is 1200 $\rm M_{\odot}yr^{-1}$,
as estimated from the infrared luminosity (e.g. Kennicutt 1998).

In this article, we adopt a standard flat cosmological model,
with $\Lambda$ = 0.73, and a Hubble constant of 71\,km\,s$^{-1}$\,Mpc$^{-1}$
(Hinshaw \etal  2009).

%--------------------------------------One column Table-----------------1-
\begin{table}[h]
      \caption[]{Definition of the sample, selected by FIR luminosity}
         \label{tab:sample}
            \begin{tabular}{l c c c c}
            \hline
            \noalign{\smallskip}
  S & Source   & RA(2000) & DEC(2000) & z \\
            \noalign{\smallskip}
            \hline
            \noalign{\smallskip}
S1& $^a$J005009.81-003900.5 & 00:50:09.8& -00:39:01 & 0.727\\
S2&  FF J0139+0115            & 01:39:27.4& +01:15:52 & 0.612\\
S3&  IRAS Z02317-0152         & 02:34:21.8& -01:39:01 & 0.645\\
S4& $^a$J014612.79+005112.4&   01:46:12.8&  00:51:12& 0.622 \\
S5&  IRAS Z02433+0110         & 02:45:55.5& +01:23:29 & 0.798\\
S8&$^a$J080430.99+360718.1 & 08:04:31.0& +36:07:18 & 0.656\\
S9&$^a$J081929.48+522345.2 & 08:19:29.5& +52:23:45 & 0.624\\
S10&$^a$J084846.34+022034.1 & 08:48:46.3& +02:20:34 & 0.627\\
S11&$^a$J091501.71+241812.1 & 09:15:01.7& +24:18:12 & 0.840\\
S12&$^a$J092527.80+001837.1 & 09:25:27.8& +00:18:37 & 0.812\\
S13&$^a$J093741.65+385752.6 & 09:37:41.6& +38:57:53 & 0.621\\
S14&  IRAS F10398+3247         & 10:42:40.8& +32:31:31 & 0.633\\
%S15&  IRAS F11038+3217         & 11:06:35.7& +32:01:46 & 0.900\\
S16&  SBS 1148+549             & 11:51:20.4& +54:37:33 & 0.975\\
S18&  3C 280                   & 12:56:57.1& +47:20:20 & 0.996\\ 
S20&  3C 289                   & 13:45:26.4& +49:46:33 & 0.967\\
S21&$^a$J144642.28+011303.0 & 14:46:42.3& +01:13:03 & 0.725\\
S22&$^a$J144953.69+444150.3 & 14:49:53.7& +44:41:50 & 0.670\\
S23&  IRAS F14537+1950         & 14:56:04.4& +19:38:46 & 0.640\\
S25&$^a$J153244.01+324246.6 & 15:32:44.0& +32:42:47 & 0.926\\
S26&$^a$J160222.38+164353.7 & 16:02:22.4& +16:43:54 & 0.672\\
S27&$^a$J160825.24+543809.8 & 16:08:25.2& +54:38:10 & 0.907\\
S28&  CFN1 078                 & 16:12:37.0& +54:28:51 & 0.910\\
S29&$^a$J161422.14+323403.6 & 16:14:22.1& +32:34:04 & 0.710\\
S30&$^a$J172507.39+370932.1 & 17:25:07.4& +37:09:32 & 0.689\\
S32&  IRAS  17548+6706         & 17:54:49.6& +67:05:57 & 0.919\\
S34&  IRAS Z21293-0154         & 21:31:53.5& -01:41:43 & 0.730\\
S35& PG 0044+030      &        00:47:05.9& +03:19:55& 0.623\\
S36& $^b$EGS70-158   &   14:21:37.5& +53:17:18& 0.930 \\
S37& $^b$EGS70-124& 14:22:39.2& +53:24:21& 0.850\\
S38& $^a$J142301.15+532419.2& 14:23:01.1& +53:24:19& 0.660\\
S39&$^a$J142348.80+533006.3& 14:23:48.8& +53:30:06& 0.770\\
S40& $^b$EGS70-067   &   14:24:30.1& +53:35:43& 0.960\\
S41& $^b$EGS70-058   &   14:25:02.8& +53:31:25& 0.690\\
S42&$^a$J143047.32+602304.4& 14:30:47.3& +60:23:04& 0.607\\
S43& $^a$J143151.84+324327.9& 14:31:51.8& +32:43:28& 0.660\\
S44& $^a$J143820.76+340234.3& 14:38:20.7& +34:02:34& 0.670\\
S45& $^a$J172228.04+601526.0& 17:22:28.0& +60:15:26& 0.742\\
S46& $^a$J171302.41+593611.1& 17:13:02.4& +59:36:11& 0.668\\
S47& $^a$J171312.08+600840.4& 17:13:12.0& +60:08:40& 0.759\\
            \noalign{\smallskip}
            \hline
           \end{tabular}
[$^{a}$] SDSS source;
[$^{b}$] [SWR2008].
Redshifts are taken from the NED data base.
\end{table}

\section{Observations}
\label{obs}

The observations were carried out with the IRAM 30m telescope
at Pico Veleta, Spain, in two periods, January and June 2011,
with a few remaining sources observed in the pool between November 2011
and February 2012.  According to their redshifts, sources were observed
in their CO(2-1), CO(3-2) or CO(4-3) lines, either with the 3mm, 2mm or 1mm receivers.
  Often only one line was observable. When possible and when the atmosphere
was favorable, we simultaneously observed two CO lines.  This was possible for seven
galaxies (see Table \ref{tab:lines}).

The broadband EMIR receivers were tuned in single sideband mode,
with a total bandwidth of 4 GHz per polarization. This covers a velocity range
of $\sim$ 12,000 \kms at 3mm and $\sim$ 8,000 \kms at 2mm.  The
observations were carried out in wobbler switching mode, with reference
positions offset by $2\arcmin$ in azimuth. Several backends were used in parallel,
the WILMA autocorrelator with  $2$~MHz channel width, covering 4$\times$4\,GHz,
and the $4$~MHz filterbanks, covering 2$\times$4\,GHz.

We spent on average three hours on each galaxy, and reached
a noise level between 0.7 and 2 mK (main beam temperature), smoothed
over  $30$~km~s$^{-1}$ channels for all sources. 
Pointing measurements were carried out every two hours on continuum sources and
the derived pointing accuracy was 3$''$ rms.  The temperature scale used is main
beam temperature $T_{\rm mb}$. 
At 3mm, 2mm and 1mm, the telescope half-power beam
width is 27$''$, 17$''$ and 10$''$, respectively. 
The main-beam efficiencies are $\eta$$_{\rm mb}=T_{\rm
  A}^*/T_{\rm mb}$=0.85, 0.79 and 0.67, respectively, 
and $S/T_{\rm mb}$ = 5.0 Jy/K for all bands. 
Spectra were reduced with the CLASS/GILDAS software, and
the spectra were smoothed up to $\sim 60$~km~s$^{-1}$ channels for the plots.
\footnote{Spectra are available in electronic form
at the CDS via anonymous ftp to cdsarc.u-strasbg.fr (130.79.128.5)
or via http://cdsweb.u-strasbg.fr/cgi-bin/qcat?J/A+A/
}

%--------------------------------------------Two column Table-2-
\begin{table*}
      \caption[]{Observed line parameters}
         \label{tab:lines}
            \begin{tabular}{l c c c c c c c c c c c}
            \hline
            \noalign{\smallskip}
Gal &Line&$\nu_{\rm obs}$&S(CO)$^{a}$ & $V^{b}$&$\Delta V_{\rm FWHM}$&L'$_{\rm CO}$/10$^{10}$&S$_{60}$&S$_{100}^{c}$&
log L$_{\rm FIR}$ & F(1.4GHz)$^{d}$ & log M$_*^f$\\
         &          & [GHz]             &  [Jy \kms]        & [\kms]      &[\kms]     & [K  \kms\, pc$^2$]&[Jy]&[Jy]& [\lsol] & [mJy]&[\msol]\\
            \noalign{\smallskip}
            \hline
            \noalign{\smallskip}

   S1  &  CO(21)  & 133.490  &   2.1  $\pm$  0.3  &   354.  $\pm$   37.  &   469.  $\pm$   73.  &    1.53 &   0.26 &   0.44 &  13.09 &   4.3 &  10.84$^g$ \\
       &  CO(43)  & 266.960  &   4.1  $\pm$  1.1  &   428.  $\pm$   96.  &   702.  $\pm$  208.  &    0.74 &  & & &  &  \\
   S2  &  CO(21)  & 143.014 &  $<$  1.2  &  &  & $<$   0.6 &   0.23 &   0.21 &  12.77 &    9.0 &  10.58 \\
   S3  &  CO(21)  & 140.145 &  $<$  1.0  &  &  & $<$   0.6 &   0.14 &   0.03 &  12.51 &   70.3 &   9.58 \\
   S4  &  CO(43)  & 284.242 &  $<$  1.0  &  &  & $<$   0.1 &   0.06 &  $<$0.12 &  12.32 &     -- &  11.55 \\
   S5  &  CO(43)  & 256.419 &  $<$  3.0  &  &  & $<$   0.7 &   0.21 &   0.40 &  13.12 &    2.0 &  10.84$^g$ \\
   S8  &  CO(21)  & 139.196 &  $<$  1.2  &  &  & $<$   0.7 &   0.03 &   0.36 &  12.58 &   73.7 &  10.71 \\
   S9  &  CO(21)  & 141.957 &  $<$  1.0  &  &  & $<$   0.5 &   0.25 &   0.03 &  12.72 &    2.6 &  10.20 \\
  S10  &  CO(21)  & 141.698  &   1.3  $\pm$  0.2  &   132.  $\pm$   27.  &   291.  $\pm$   64.  &    0.68 &   0.15 &   0.03 &  12.51 &   1.1 &  10.59 \\
  S11  &  CO(43)  & 250.566 &  $<$  4.1  &  &  & $<$   1.0 &   0.25 &   0.03 &  13.03 &    9.3 &  11.01$^h$ \\
  S12  &  CO(32)  & 190.837  &   4.8  $\pm$  1.4  &    20.  $\pm$   36.  &   271.  $\pm$  102.  &    1.93 &   0.15 &   0.03 &  12.79 &    -- &  10.42 \\
  S13  &  CO(21)  & 142.220 &  $<$  0.8  &  &  & $<$   0.4 &   0.22 &   0.32 &  12.83 &    0.7 &  11.43 \\
  S14  &  CO(21)  & 141.175  &   1.1  $\pm$  0.2  &     1.  $\pm$   15.  &   122.  $\pm$   29.  &    0.61 &   0.22 &   0.54 &  12.94 &   6.3 &  10.62 \\
       &  CO(43)  & 282.327  &   3.7  $\pm$  0.9  &   -82.  $\pm$   65.  &   493.  $\pm$   99.  &    0.49 & & & & &  \\
% S15  &  CO(43)  & 242.653 &  $<$  1.8  &  &  & $<$   0.5 &   0.23 &   0.70 &  13.39 &   13.7 &  10.72 \\
  S16  &  CO(21)  & 116.704 &  $<$  3.2  &  &  & $<$   4.2 &   0.20 &   0.41 &  13.33 &    4.5 &  12.66$^{gh}$ \\
       &  CO(43)  & 233.389  &  10.3  $\pm$  1.6  &   521.  $\pm$   31.  &   483.  $\pm$  103.  &    3.36 & & & & &  \\
  S18  &  CO(43)  & 230.982 &  $<$  2.0  &  &  & $<$   0.7 &   0.12 &   0.06 &  12.95 & 5100.0 &  10.33 \\
  S20  &  CO(43)  & 234.340 &  $<$  1.8  &  &  & $<$   0.6 &   0.10 &   0.03 &  12.82 & 2400.0 &  11.66$^g$ \\
  S21  &  CO(21)  & 133.614 &  $<$  0.8  &  &  & $<$   0.6 &   0.08 &   0.15 &  12.60 &    5.9 &  11.54 \\
  S22  &  CO(21)  & 138.047  &   1.8  $\pm$  0.4  &  -656.  $\pm$   25.  &   249.  $\pm$   57.  &    1.11 &   0.19 &   0.03 &  12.68 &  11.0 &  10.04 \\
       &  CO(43)  & 276.072  &   6.0  $\pm$  0.9  &  -685.  $\pm$   16.  &   214.  $\pm$   33.  &    0.91 & & & & &  \\
  S23  &  CO(21)  & 140.572 &  $<$  1.0  &  &  & $<$   0.6 &   0.28 &   0.03 &  12.79 &     -- &  10.29 \\
  S25  &  CO(43)  & 239.415 &  $<$  3.0  &  &  & $<$   0.9 &   0.28 &   0.50 &  13.40 &    5.9 &  11.36$^g$ \\
  S26  &  CO(21)  & 137.882 &  $<$  1.2  &  &  & $<$   0.7 &   0.15 &   0.03 &  12.59 &    1.9 &  11.85$^h$ \\
  S27  &  CO(32)  & 181.330  &   1.5  $\pm$  0.3  &  -128.  $\pm$   28.  &   211.  $\pm$   60.  &    0.77 &   0.03 &   0.26 &  12.82 &   0.2 &  11.25 \\
  S28  &  CO(43)  & 241.383 &  $<$  2.2  &  &  & $<$   0.6 &   0.03 &   0.24 &  12.80 &     -- &  10.14 \\
  S29  &  CO(21)  & 134.818  &   4.4  $\pm$  0.5  &    68.  $\pm$   18.  &   373.  $\pm$   48.  &    3.02 &   0.17 &   0.21 &  12.84 &   1.2 &  11.39$^g$ \\
       &  CO(43)  & 269.614  &   8.4  $\pm$  1.6  &    40.  $\pm$   38.  &   421.  $\pm$   85.  &    1.43 & & & & & \\
  S30  &  CO(21)  & 136.494  &   2.4  $\pm$  0.4  &  -133.  $\pm$   21.  &   267.  $\pm$   43.  &    1.52 &   0.24 &   0.23 &  12.92 &   2.3 &  11.04$^g$ \\
       &  CO(43)  & 272.967  &   1.3  $\pm$  0.4  &   -32.  $\pm$    6.  &    44.  $\pm$   26.  &    0.21 & & & & & \\
  S32  &  CO(43)  & 240.251 &  $<$  1.8  &  &  & $<$   0.5 &   0.40 &   0.47 &  13.48 &     -- &  10.72 \\
  S34  &  CO(21)  & 133.259  &   3.6  $\pm$  0.5  &   -60.  $\pm$   41.  &   569.  $\pm$   96.  &    2.58 &   0.19 &   0.56 &  13.07 &   2.7 &  10.84$^g$ \\
       &  CO(43)  & 266.498  &   4.8  $\pm$  1.1  &   -34.  $\pm$   52.  &   414.  $\pm$  104.  &    0.87 & & & & & \\
  S35  &  CO(21)  & 142.022  &   1.8  $\pm$  0.5  &    93.  $\pm$   19.  &   182.  $\pm$   70.  &    0.93 &   0.07 &   0.16 &  12.41 & 166.4 &  11.79$^h$ \\
  S36  &  CO(43)  & 238.881 &  $<$  3.6  &  &  & $<$   1.1 &   0.03$^{e}$ &   0.14$^{e}$ &  12.66 &     -- &  11.62 \\
  S37  &  CO(43)  & 249.211 &  $<$  2.8  &  &  & $<$   0.7 &   0.01$^{e}$ &   0.06$^{e}$ &  12.21 &     -- &  11.29 \\
  S38  &  CO(21)  & 138.878 &  $<$  1.4  &  &  & $<$   0.8 &   0.02$^{e}$ &   0.08$^{e}$ &  12.04 &     --&  10.83 \\
  S39  &  CO(21)  & 130.247 &  $<$  1.2  &  &  & $<$   1.0 &   0.01$^{e}$ &   0.10$^{e}$ &  12.25 &     --&  12.19$^h$ \\
  S40  &  CO(43)  & 235.225 &  $<$  1.8  &  &  & $<$   0.6 &   0.01$^{e}$&   0.09$^{e}$ &  12.42 &     --&  11.69 \\
  S41  &  CO(21)  & 136.413 &  $<$  1.0  &  &  & $<$   0.7 &   0.01$^{e}$&   0.12$^{e}$ &  12.15 &     --&  11.77 \\
  S42  &  CO(21)  & 143.472  &   2.3  $\pm$  0.4  &   -98.  $\pm$   93.  &  1001.  $\pm$  244.  &    1.12 &   0.06 &   0.20 &  12.41 &  15.0 &  11.86 \\
  S43  &  CO(21)  & 138.878  &   4.7  $\pm$  0.6  &   560.  $\pm$   23.  &   385.  $\pm$   49.  &    2.75 &   0.05$^{e}$&   0.14$^{e}$ &  12.38 &    -- &  11.05$^g$ \\
  S44  &  CO(21)  & 138.047  &   5.9  $\pm$  1.0  &    60.  $\pm$   36.  &   415.  $\pm$   72.  &    3.59 &   0.07$^{e}$ &   0.25$^{e}$ &  12.57 &    -- &  11.38$^g$ \\
  S45  &  CO(21)  & 132.345 &  $<$  1.8  &  &  & $<$   1.4 &   0.04$^{e}$&  $<$0.15$^{e}$ &  12.47 &    1.2 &  11.67 \\
  S46  &  CO(21)  & 138.212 &  $<$  1.4  &  &  & $<$   0.9 &   0.04$^{e}$&  $<$0.16$^{e}$ &  12.39 &    4.6 &  10.72 \\
  S47  &  CO(21)  & 131.030  &   3.1  $\pm$  0.7  &    46.  $\pm$   46.  &   437.  $\pm$  153.  &    2.45 &   0.03$^{e}$ &  $<$0.11$^{e}$ &  12.34 &   0.5 &  11.31 \\
            \noalign{\smallskip}
            \hline
           \end{tabular}
\begin{list}{}{}
\item[] Quoted errors are statistical errors from Gaussian
            fits. The systematic calibration uncertainty is 10\%. 
 \item[] The derivation of L$_{\rm FIR}$ is described in Sec. \ref{sample}, and log SFR(\msol/yr) = log L$_{\rm FIR}$(\lsol) - 9.76 (cf Sec.\ref{CO-mass}).
\item[$^{a}$] The upper limits are at 3$\sigma$ with an assumed $\Delta$V = 300 \kms, except for S16, where $\Delta$V is known.
\item[$^{b}$] The velocity is relative to the optical redshift given in Table \ref{tab:sample}.  
\item[$^{c}$] The 60 and 100\mic\, fluxes are from NED (http://nedwww.ipac.caltech.edu/) or Stanford \etal (2000)
\item[$^{d}$] From the FIRST  catalog (http://sundog.stsci.edu/). Errors are typically 0.14 mJy
\item[$^{e}$] For these objects, the Spitzer 70 and 160\mics fluxes replace the IRAS 60 and 100\mics ones.
\item[$^{f}$] Stellar masses were obtained through SED fitting (see Sec. \ref{GFR}), with SDSS fluxes 
for all galaxies, and in addition, with Spitzer-IRAC data for galaxies with [$^{g}$] and 2MASS for galaxies 
with [$^{h}$] in the last column.
\end{list}
\end{table*}

\section{Results}
\label{res}

\subsection{CO detection in z=0.6-1.0 ULIRGs}
\label{COdet}

Figures \ref{fig:spec1}, \ref{fig:spec2} and \ref{fig:spec3}
display the CO-detected sources, in their lower-J CO spectrum. For three galaxies,
 S22, S29 and S34,
their higher-J lines are also plotted.  Table \ref{tab:lines} reports
all line parameters, and also the upper limits for the non detections.
Integrated signals and velocity widths have been computed from Gaussian fits.
These also give the central velocities, with respect to the optical
redshift of Table \ref{tab:sample}. There are some offsets between
the CO and optical velocities, always lower than 700\kms. These offsets might
be due to the accuracy of the optical determination, or be intrinsic due to
an outflow of ionized gas.
The upper limits are computed at 3$\sigma$,
assuming a common line width of 300\kms and getting
the rms of the signal over 300\kms. Lines are considered detected 
when the integrated signal is more than 3$\sigma$. 
All spectra above 3$\sigma$ are shown in the figures.

The detection rate of 38\% in this z=0.6--1 sample is lower than the value of 60\%
found in the lower-z counterpart at z=0.2-0.6 (Paper II); however, this can already be explained
by the reduced sensitivity for the more distant objects. Indeed, the 
lines detected are now CO(2-1) instead of CO(1-0), and for excited point sources
the flux could be up to about four times larger; however, the average luminosity distance
at 0.6 $<$ z $<$ 1.0 weakens the signal by about a factor 5.5 with respect
to the previous closer sample, and this outweighs the gain by climbing up
 the CO ladder. 
 At  0.6 $<$ z $<$ 1.0, the average angular distance is 1550 Mpc; i.e., the beam
of 17$''$  corresponds to 127 kpc, so the sources can all be considered as point-like
at this resolution.

  The line widths detected are compatible with what is
expected from massive galaxies of ULIRG type, and quite comparable
to what was found in our lower-z sample.
Their average is $\Delta$V$_{\rm FWHM}$= 370 \kms, compared
to 348\kms\ in Paper II. Some galaxies have somewhat different widths
in their two detected CO lines, but this is likely due to the noise intrinsic to the data.

\subsection{CO luminosity and \hh\, mass}
\label{CO-mass}

Since we have not observed the fundamental CO(1-0) line, which 
is a direct measure of the total \hh\, mass, but mostly CO(2-1), it
is interesting to compute L'$_{CO}$, the special unit CO luminosity,
through integrating the CO intensity over the velocity profile.
  This luminosity, expressed in units of K \kms pc$^2$,
 will give the same value irrespective of  J, if the
CO lines are saturated and have the same brightness temperature.

This CO luminosity is given by $$L'_{CO} =
23.5 I_{CO} \Omega_B {{D_L^2}\over {(1+z)^3}} \hskip6pt \rm{K\hskip3pt
  km \hskip3pt s^{-1}\hskip3pt pc^2}$$
where  $I_{CO}$ is the intensity in K \kms, $\Omega_B$ the area of
the main beam in square arcseconds, and $D_L$ the luminosity
distance in Mpc.  As mentioned above, all sources can be considered to be unresolved,
since our beam is typically 130kpc in size.
We here assume a ratio of 1 between the
CO(2-1) and CO(1-0) luminosities (or brightness temperatures),
 as expected for a warm optically thick, and thermally excited medium. 
In Paper II, we found some sources with lower excitation,
but the excitation appears higher in the present sample,
as discussed in section \ref{excit}. 
In any case, our hypothesis of a CO(2-1)/CO(1-0) ratio equal
to unity can only underestimate the CO(1-0) luminosity and thus the
molecular masses. Under this assumption,
we compute H$_2$ masses using
M$_{\rm H_2} = \alpha$ L'$_{\rm CO}$, 
with $\alpha=0.8$ M$_\odot$ (K \kms\, pc$^2$)$^{-1}$,
the appropriate factor for ULIRGs. The molecular gas masses are
listed in Table~\ref{tab:h2}. 
 The choice of a common conversion factor might not be realistic,
but it is a first approximation before knowing more
physical quantities about each source, and adapting an individual
factor for each, based on observation of more CO lines and of 
the gas spatial extent.
  In the following, CO luminosities are often used instead
of H$_2$ masses, to remind the reader of this uncertainty.

The average CO luminosity for the 15 galaxies detected
is L'$_{\rm CO}$ = 1.8 10$^{10}$ K \kms\, pc$^2$, corresponding to an 
average \hh\, mass of  1.5 10$^{10}$ \msol.
The star formation efficiency (SFE), also listed in Table~\ref{tab:h2},
is defined as  L$_{\rm FIR}$/M(\hh) in \lsol/\msol. The SFR
is related to the FIR luminosity, where we adopt the relation 
SFR= L$_{\rm FIR}$ /(5.8 10$^9$L$_{\odot}$) compiled by Kennicutt (1998).
The gas consumption time scale can then be derived as 
$\tau$ = 5.8/SFE  Gyr, where SFE is in units of \lsol/\msol.

%--------------------------------------One column Table----------------3--
\begin{table}[h]
      \caption[]{ Molecular gas mass, star formation efficiency, dust temperature
and dust mass, half-light radius and galaxy type of the sample}
         \label{tab:h2}
            \begin{tabular}{l c c c c c c}
            \hline
            \noalign{\smallskip}
  S & M(\hh)   & SFE  & T$_d$ & M$_d$ & R$_{1/2}$ & Type\\
&  10$^9$ M$_\odot$ &  \lsol/\msol\, &  K   &  10$^8$ M$_\odot$ & kpc & \\
            \noalign{\smallskip}
            \hline
            \noalign{\smallskip}
  S 1 &   12.3   &   1003.   &  58.7  &   1.4 & 1.1$^a$ & S2,Q,wi  \\
  S 2 &  $<$4.9  &   $>$1205. &  70.1  &     0.3 &  2.9$^a$ &wi \\
  S 3 &   $<$4.5 &    $>$713. &  66.5  &     0.2 &  0.5$^a$ & \\
  S 4 &    $<$1.1 &   $>$1978. &  $>$ 52.0  &    $<$0.5 & 3.9 &S2,Q,si  \\
  S 5 &    $<$5.3 &   $>$2504. &  58.6  &     1.6 &  3.1$^a$ &si \\
  S 8 &    $<$5.6 &    $>$674. &  33.4  &    20.7 & 7.3 &S2,Q  \\
  S 9 &    $<$4.2 &   $>$1238. &  68.2  &     0.3 & 4.2 &S1  \\
  S10 &    5.4 &    597. &  67.7  &     0.2 & 8.5 &S2  \\
  S11 &    $<$7.8 &   $>$1375. &  76.0  &     0.4 & 1.1$^a$ &Q,si  \\
  S12 &   15.4 &    400. &  75.4  &     0.2 & 5.0 &S2  \\
  S13 &    $<$3.4 &   $>$2012. &  58.2  &     0.8 &  2.2$^a$ & \\
  S14 &    4.9 &   1776. &  49.0  &     3.0 &  0.9$^a$ &si \\
%  S15 &    $<$4.0 &   $>$6088. &  53.4  &     5.5 & 3.1$^a$ &  \\
  S16 &   21.0 &    1016. &  62.4  &     1.8 & 2.9$^a$ &Q  \\
  S18 &    $<$5.5 &   $>$1624. &  81.0  &     0.2 &  3.4$^a$ & \\
  S20 &    $<$4.7 &   $>$1419. &  69.0  &     0.4 &  3.3$^a$ &si \\
  S21 &    $<$4.6 &    $>$861. &  56.5  &     0.6 & 6.4 &S2,Q  \\
  S22 &    8.8 &    541. &  77.2  &     0.2 & 1.5$^a$ &S2  \\
  S23 &    $<$4.5 &  $>$ 1381. &  72.2  &     0.3 &  3.5 & \\
  S25 &    $<$7.1 &   $>$3536. &  64.2  &     1.8 & 1.3$^a$ &S2,si  \\
  S26 &    $<$5.9 &    $>$656. &  70.2  &     0.2 &  6.2 &wi \\
  S27 &    6.2 &   1070. &  41.3  &     9.4 &  7.9 & \\
  S28 &    $<$5.0 &   $>$1254. &  42.2  &     7.7 &  -- & \\
  S29 &   24.1 &    286. &  65.4  &     0.4 &  1.6$^a$ &si \\
  S30 &   12.2 &    683. &  71.8  &     0.3 &  1.1$^a$ &wi \\
  S32 &    $<$4.2 &   $>$7190. &  74.8  &    0.9 &  4.0 & \\
  S34 &   20.7 &    569. &  49.1  &     4.2 &  7.8$^a$ &si \\
  S35 &    7.4 &    346. &  50.0  &     0.8 & 2.8$^a$ &S2,Q,wi  \\
  S36 &    $<$8.6 &   $>$ 531. &  35.1  &    18.6 &  6.4 & \\
  S37 &    $<$5.6 &   $>$ 284. &  34.2  &     7.0 &  -- & \\
  S38 &    $<$6.7 &   $>$ 161. &  30.4  &     9.1 &  2.2 &wi \\
  S39 &    $<$7.8 &    $>$222. &  29.9  &    23.5 &  11. &wi \\
  S40 &    $<$4.6 &    $>$574. &  33.1  &    20.5 &  -- & \\
  S41 &    $<$5.2 &   $>$265. &  26.3  &    55.6 &  -- & \\
  S42 &    8.9 &    287. &  44.0  &     1.7 & 7.8 &Q,si  \\
  S43 &   22.0 &    109. &  34.1  &     6.9 &  5.6 &wi \\
  S44 &   28.7 &    129. &  31.9  &    21.1 &  5.6 &wi \\
  S45 &   $<$10.9 &    $>$265. & $>$32.8 &   $<$14.8 &  4.0 &si \\
  S46 &    $<$6.8 &    $>$343. & $>$31.3  &   $<$15.8 &  5.5 & \\
  S47 &   19.6 &    112. & $>$33.1  &   $<$10.9 &  3.1$^a$ &si \\
            \noalign{\smallskip}
            \hline
           \end{tabular}
\\ M(\hh) and SFE are defined in Sec. \ref{SFE}, T$_d$ and M$_d$ in Sec. \ref{CO-FIR}
\\ R$_{1/2}$ is defined from red images either from SDSS, or HST for the galaxies with index [$^a$]
\\The latter images were used to determine the morphological class: 
wi, si= weak \& strong interaction. The nuclear activity type is derived from NED:
S1, S2= Seyfert 1 \& 2, Q= QSO 
\end{table}

\subsection{Molecular gas excitation}
\label{excit}

Seven sources have been observed in two CO lines,
CO(2-1) and CO(4-3), and  six have been detected in both,
as listed in Table \ref{tab:COex}.
One has an upper limit, in the CO(2-1) line.
To compare the two lines, and derive the excitation,
we computed both the total and the peak flux ratio 
 S$_{43}$/S$_{21}$ 
since the CO(4-3) and CO(2-1) lines have sometimes different measured linewidths,
which we attribute  to noise.  The corresponding ratios
between the peak brightness temperatures are also displayed in Table
\ref{tab:COex}, to allow for easy comparison with the predictions of excitation models.
 The temperature ratio is corrected for the different beam sizes, assuming
that the sources are unresolved in all our observations.

The excitation essentially depends on two parameters, the \hh\, volume
density, and the kinetic temperature.  In ultraluminous objects,
the column densities of the molecular gas are high enough (N(\hh) $>$ 10$^{24}$ cm$^{-2}$)
that the CO lines are always optically thick in the low-J lines. This result was reached
through mapping the gas content of local ULIRGs, where the gas is very concentrated (e.g. Solomon \etal 1997).
Depending on the velocity width of the lines, the CO column density per unit velocity width (km/s)
is higher than 10$^{17}$ cm$^{-2}$. To constrain the kinetic temperature
of the gas, 
we computed the dust temperature deduced from the far-infrared fluxes 
(see Table~\ref{tab:h2}),
 assuming $\kappa_\nu\propto\nu^{\beta}$, where
$\kappa_\nu$ is the mass opacity of the dust at frequency $\nu$, and
$\beta$ = 1.5. The average dust temperature for our  0.6$<$z$<$1.0 sample
 is $54\pm5$~K,  higher than the average dust temperature for the 0.2$<$z$<$0.6 sample
 of $46\pm5$~K (Paper II). 
This is also higher than what is obtained for local starburst galaxies, which have dust
temperatures $\approx40$~K (e.g. Sanders \& Mirabel 1996,  Elbaz \etal 2010).
Yang \etal (2007) have observed seven of our objects at 350\mics and derive more precise 
temperatures, which are very similar to the values computed above.
 We note a clear increase in the dust temperatures with respect to our
0.2 $<$ z $<$ 0.6 sample in Paper II. This could be caused by selection effects,
and also be due to the frequency used to measure the temperature. The 60 and 100\mics bands
correspond to rest-frame 45 and 76\mics at z$\sim$ 0.32, and to rest-frame 35 and 58\mics at  z$\sim$ 0.72,
the median redshift of the two samples.

As in Paper II, we assume that
 the gas is predominantly heated by collisions with the dust grains. 
Indeed, there could be small regions that are heated by
the UV photons of young massive stars or by shocks, but
our beam encompasses kpc scales, and these would be spatially
diluted.
 The gas kinetic temperature should be at most equal to the dust temperature,
and this is the constraint used in our LVG modeling.
We compare the CO-based molecular masses and dust masses in section 
\ref{CO-FIR}.

\begin{figure*}
\centering
\includegraphics[angle=-90,width=16cm]{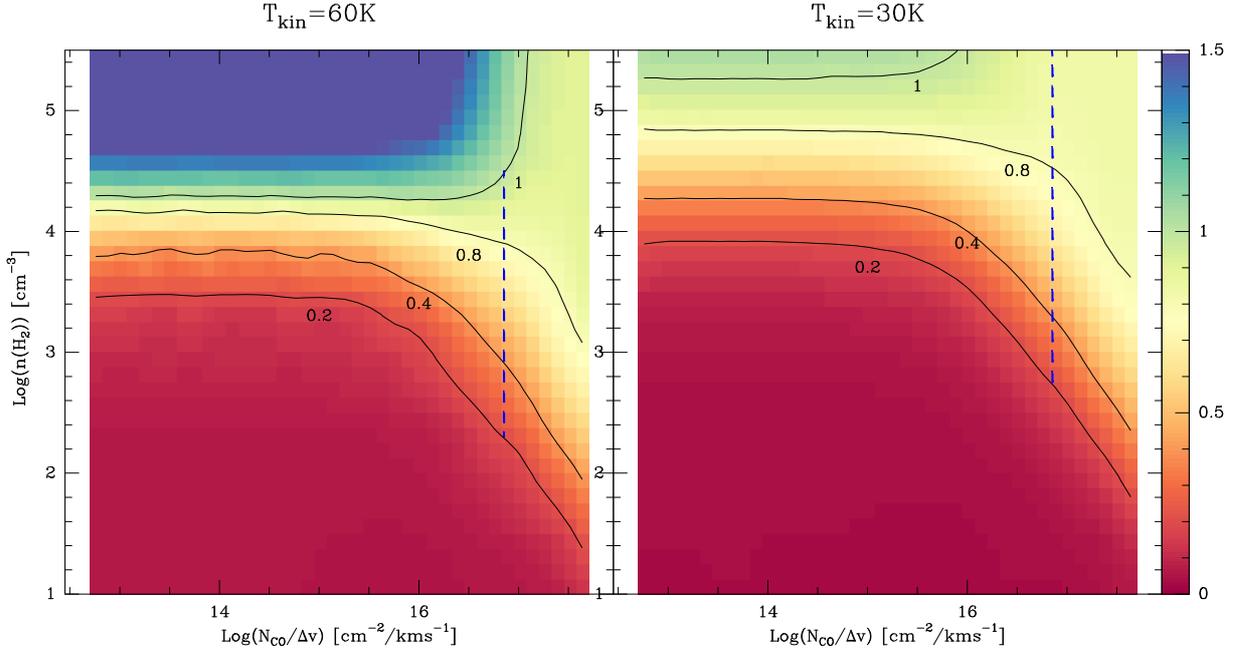}
\caption{Peak T$_{\rm b}$ ratio between the CO(4-3) and CO(2-1) lines
versus \hh\, density, and the CO column density per unit velocity
width (N$_{\rm CO}/\Delta V$) for two values of the  kinetic temperature:
T$_k$ = 60K, the dust temperature (left), and T$_k$ = 30K (right).
The black contours are underlining the values
obtained in the data. 
 In both panels, the vertical line corresponding to (N$_{\rm CO}/\Delta V$)=
7$\times$10$^{16}$ cm$^{-2}$/(\kms), value chosen in Table \ref{tab:COex}, has been
emphasized with blue dashes.
The predictions come from the LVG models that are part of
the Radex code (van der Tak 2007).} 
\label{fig:lvg}
\end{figure*}

 Using the Radex code (van der Tak \etal  2007), we computed
the predicted main beam temperature ratio between the 
CO(4-3) and CO(2-1) lines, for several kinetic temperatures
and as a function of \hh\, densities, and CO column densities.
Figure \ref{fig:lvg} shows these predictions for T$_k$=60\,K and 30\,K. 
The black contours delineate the range of observed values.
 In Table  \ref{tab:COex} we list the derived values for the n(\hh) densities,
for two values of the kinetic temperatures (60 and 30K), and for a fixed
column density per velocity width.

We adopted a column density of N(CO)/$\Delta V$
of 7$\times$10$^{16}$ cm$^{-2}$/(\kms), which corresponds to
what is expected if the 
high molecular gas masses derived in Table \ref{tab:lines}
are highly concentrated in the few central kpc. 
For M(\hh) = 3 10$^{10}$ M$_\odot$, a typical CO abundance of 
CO/\hh\, = 10$^{-4}$, and a linewidth of 300 \kms,
this column density corresponds to a homogenous 
disk of 3 kpc in size. This average column density would
be a lower limit, if the gas inside 3kpc is clumpy.

The excitation of the CO gas is in general higher than in 
our lower-z sample.  In Paper II, the CO(3-2)/CO(1-0) ratio was
found to be $\sim$ 3 times lower than the CO(4-3)/CO(2-1) ratio derived here, and
the \hh\, volumic densities were on the order of 100,
while they are $\sim$ 10$^3$ cm$^{-3}$ in the present analysis.
In four out of the seven sources where we have excitation constraints
through measurements of two lines,
the CO emission ladder should be populated well 
above the J=4 line. In the three remaining sources, the excitation
is typical of ``normal'' or weakly interacting galaxies, like the Milky Way
or the Antennae (e.g. Weiss \etal  2007). 

We note that these conclusions on the excitation are preliminary. 
Observations of several lines for more sources are required,
as there are large variations from
source to source. In addition, the spatial extent of the emission needs to be obtained
through interferometric measurements.

\subsection{CO luminosity and redshift}
\label{CO-z}

\begin{figure}[h!]
\centering
\includegraphics[angle=0,width=8cm]{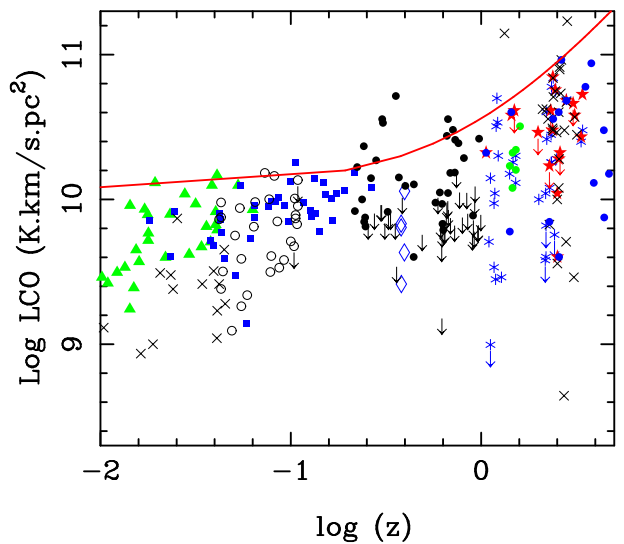}
\caption{Measured CO luminosities, corrected for
amplification when known, but not for gas excitation,
as a function of redshift.  We compare 
our points (filled black circles, and arrows as upper limits) 
with a compilation of high-z molecular gas surveys, and local ones:
green filled triangles are from Gao \& Solomon (2004),
filled blue squares from Solomon \etal (1997),
open circles from Chung \etal (2009),
open blue diamonds from Geach \etal (2009, 2011),
black crosses from Iono \etal (2009),
red stars from Greve \etal (2005),
green filled circles from Daddi \etal (2010),
blue asterisks from Genzel \etal (2010), and  blue filled circles 
from Solomon \& vanden Bout (2005). 
For illustration purposes only,
the red curve is the power law in (1+z)$^{1.6}$ for $\Omega_\hh$/$\Omega_\hi$
proposed by Obreschkow \& Rawlings (2009).}
\label{fig:CO-z}
\end{figure}

As in Paper II, we can now investigate the evolution of 
the molecular gas content of galaxies with redshift.
The CO luminosity of our sample objects, selected to be bright with high
masses, is now compared to data reported in the literature at different redshifts, 
also selected with high masses, in Figure \ref{fig:CO-z}.  
 The local sample is the compilation of 65 infrared galaxies by Gao \& Solomon (2004),
including nine ULIRGs (L$_{\rm FIR}>$10$^{12}$\lsol), 
22 luminous infrared galaxies LIGs (10$^{11}$\lsol$<$L$_{\rm FIR}<$10$^{12}$\lsol), 
and 34 large spiral galaxies. We include 37  ULIRGs from the study
by Solomon \etal (1997) up to z=0.3, and 29 ULIRGs from Chung \etal (2009) up to z=0.1.
At intermediate redshift of 0.4, we include five massive star-forming galaxies
selected at 24\mic, and detected by Geach \etal (2009, 2011).
They are compared with the compilation
by Iono \etal (2009) of 43 low and high-redshift U/LIRGs,
submm selected galaxies (SMGs), quasars, and Lyman Break Galaxies (LBGs),
19 high-z SMG from Greve \etal (2005), massive star-forming galaxies
at high-z from Daddi \etal (2010), Genzel \etal (2010) and Solomon \& van den Bout (2005).
Our full sample (filled black circles, Paper II and this work) now
fills in the CO redshift desert, between z=0.2 and 1. 
The average CO luminosity over our 33 detected galaxies
is  L'$_{\rm CO}$ = 1.9$\times$10$^{10}$ K \kms\, pc$^2$,
while the average over the local star-forming galaxies
(Gao \& Solomon 2004) is  0.36$\times$10$^{10}$ K \kms\, pc$^2$,
a factor 5 lower.
It has been already established that galaxies have more molecular gas
at high redshift, z $\sim$ 1-2 (Tacconi \etal  2010, Daddi \etal 2010), 
 while the abundance of HI is not
thought to have varied significantly since z = 1.5 (e.g. Obreschkow \& Rawlings 2009).
Now it is possible to determine that the rise is gradual, and it already starts
at z=0.2-0.3.
This trend is also discussed for the gas fraction (M$_{gas}$/(M$_{gas}$+M$_{*}$)
in Section \ref{GFR}.

Obviously, such a trend is only indicative and should be confirmed by the
study of a large number of galaxies in the main sequence of star formation, 
and by sampling a larger part of the luminosity function. Here we focus on
the most luminous objects, and the evolution we observed is only that of 
most actively starforming galaxies.
 We expect that this evolution in L'$_{CO}$ probably underestimates the evolution
in molecular mass: at higher z, the CO luminosity has been measured
in higher-J lines, and because of likely subthermal excitation, those lines
must have lower luminosity than the CO(1-0) line. 

Theoretically, it is expected  that galaxies are richer in molecular gas
at high redshift. Indeed, galaxies at a given mass should be smaller
(Somerville \etal 2008), which is confirmed by observations (e.g. Nagy \etal 2011).
As a consequence, the gas component was more compact and thus denser, and the 
higher gas pressure transforms the HI into a molecular
gas phase, increasing the \hh/\hi\, ratio. Since the abundance of HI is not
predicted to vary significantly since z=1.5, the  \hh/\hi\,  ratio reflects
the evolution of the molecular gas content, traced by CO luminosity.
Obreschkow \etal  (2009) and Obreschkow \& Rawlings (2009)
followed the \hh/\hi\, ratio statistically
over 30 million simulated galaxies, and predicted its cosmic decline as
$\Omega_\hh$/$\Omega_\hi \propto (1+z)^{1.6}$.
 To guide the eye, we have reproduced this behavior through a red line
in Figure \ref{fig:CO-z}. The amplitude of variation in 
the L'$_{CO}$ envelope appears to approximately follow the line.

\subsection{Correlation between FIR and CO luminosities}
\label{CO-FIR}

\begin{figure}[h!]
\centering
\includegraphics[angle=0,width=8cm]{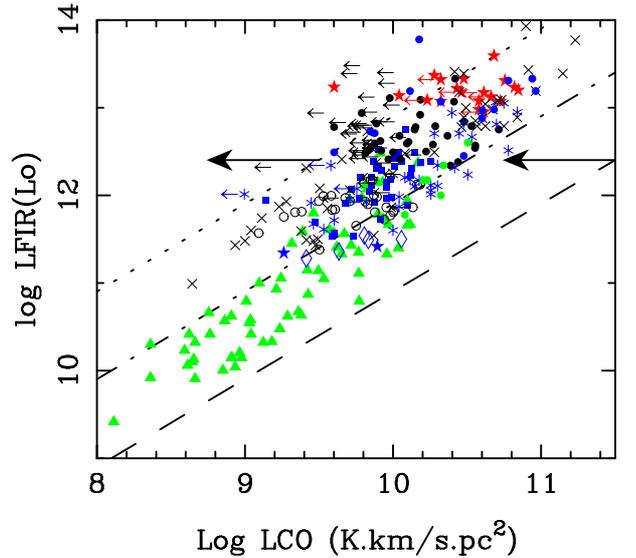}
\caption{Correlation between FIR and CO luminosities, for 
our sample (filled black circles, and arrows for upper limits) 
and other galaxies from the literature (same symbols
as in Fig \ref{fig:CO-z}). The 3 lines are for L$_{\rm FIR}$/M(\hh)=10,
100 and 1000 \lsol/\msol\, from bottom to top, assuming a conversion factor
$\alpha$ = 0.8  M$_\odot$ (K \kms\, pc$^2$)$^{-1}$. 
The same lines correspond to gas depletion time scales of 580 (bottom), 58 (middle) 
and 5.8 Myr (top).
 The arrows indicate the direction and amount by which these lines would move
if the Milky Way CO-to-\hh\, conversion factor was used.
}
\label{fig:FIR-CO}
\end{figure}

It is now well-known that a good correlation exists between
 FIR and CO luminosities (e.g. Young \& Scoville 1991). It holds
for star-forming galaxies, either in the main sequence or
in the starbursting phase, and even for quasars, implying
that their FIR luminosity is dominated by star formation
(Iono \etal 2009, Xia \etal 2012). The correlation 
is slightly non linear, and Xia \etal (2012) found a power law
of slope 1.4 as their best-fit, which is confirmed in Figure \ref{fig:FIR-CO}: this means
that ultra-luminous sources are forming stars at higher efficiency,
and with a depletion time as low as $\sim$ 10 Myr. 

In Figure \ref{fig:FIR-CO}, we draw three lines corresponding
to three orders of magnitude of the ratio  L$_{\rm FIR}$/M(\hh), if
the CO-to-\hh\, conversion factor adapted for ULIRG is used. 
Within this hypothesis, all galaxies in our sample are located above 
the curve  L$_{\rm FIR}$/M(\hh)=100 \lsol/\msol\,
(corresponding to a consumption time scale of $\tau$ =58 Myr).
However, all detected galaxies at high z in the literature are also
in this regime, including the main-sequence star-forming galaxies 
at high redshift studied by Genzel \etal (2010) or Daddi \etal (2010).
In those objects, gas excitation and spatial extent argue in favor
of a more ``normal'' conversion factor, as for the Milky Way. 
If we adopt a MW-like conversion factor, the \hh\, mass
then moves to the right by a factor 5.75 (or 0.76 in log), and these
objects will fall in the range of L$_{\rm FIR}$/M(\hh) between 10 and 100 \lsol/\msol\,
with a depletion time scale approaching the Gyr. In such conditions, 
star formation could be sustained continuously, given the  frequent
cold gas accretion from the intergalactic medium.

To better constrain the correct conversion factor to 
apply to each object, we derive the dust mass from the far-infrared
measurements, and compare it with the gas mass.
Given the already derived dust temperature T$_d$ and the observed 100~\mic\,
flux S$_{100}$, we can estimate the dust mass as
$$
\begin{array}{lcl}
 {\rm M}_d & = & 4.8\times10^{-11}\, {{S_{\nu o}\,D_{\rm Mpc}^{\,2}}
 \over {(1+z)\kappa_{\nu r}\,B_{\nu r}(T_d)}}\ \msol \\
             &  =&  5(1+z)^{-(4+\beta)}\,S_{100\mu{\rm m}}\,D_{\rm Mpc}^{\,2}\,
  \left\{\exp(144(1+z)/T_d) - 1 \right\}\ \msol\, \\
\end{array}
$$ 
where $S_{\nu o}$ is the observed FIR flux measured in Jy, $D_{\rm Mpc}^{\,2}$ the luminosity
distance in Mpc, $B_{\nu r}$ the Planck function at the rest
frequency $\nu r = \nu o (1+z)$, and we use a mass
opacity coefficient of $25$~cm$^{2}$~g$^{-1}$ at rest frame 100~\mics 
(Hildebrand 1983,  Dunne \etal  2000, Draine 2003), with a frequency dependence of $\beta$=1.5.
Estimated dust masses are displayed in Table \ref{tab:h2}.
 If we adopt the low conversion factor of $\alpha$ = 0.8  M$_\odot$ (K \kms\, pc$^2$)$^{-1}$,
the average gas-to-dust mass ratio is 206 for the detected galaxies.
 The gas-to-dust mass ratio would increase up to 1200 if the standard (MW)
conversion factor is used. 
Since the Milky Way gas-to-dust ratio is 150, which is also a value typically found in
nearby galaxies in the SINGS sample (Draine \etal 2007), while values higher than 1000 are
only found in elliptical galaxies, or in low-metallicity dwarfs 
(Wiklind \etal  1995, Leroy \etal 2011), we conclude that the ULIRG conversion factor
is not far from being adequate for our sample.
 We note that for the high stellar masses of the sample galaxies, when using the mass metallicity relation,
and its evolution by a factor 2-3 with redshift up to z=0.75 (Moustakas \etal 2012), 
we could expect an average gas-to-dust mass ratio of no more than 500 for our sample.

%--------------------------------------One column Table-------very small (4)
\begin{table}[h]
\centering
      \caption[]{CO gas excitation}
         \label{tab:COex}
            \begin{tabular}{l c c c c c}
            \hline
            \noalign{\smallskip}
  S & S$_{43}$/S$_{21}$ &  S$_{43}$/S$_{21}$ &  T$_{\rm b43}$/T$_{\rm b21}$      & n(\hh) &   n(\hh) \\
    &    [total]      &    [peak]     &   [peak]      & [cm$^{-3}$] &    [cm$^{-3}$] \\
    &       &                        &                       &             60K         &        30K   \\
            \noalign{\smallskip}
            \hline
            \noalign{\smallskip}
  S1  &2.0 &  1.3  $\pm$ 0.3  &   0.3 $\pm$ 0.07   &  4.E2 & 1.2E3 \\
  S14 &3.4 &  1.0  $\pm$ 0.3  &     0.25 $\pm$ 0.07  &  3.E2 & 1.E3 \\
  S16 &$>$3.2 &$>$3.0 & $>$ 0.75 &   $>$6.E3 &    $>$2.E4 \\
  S22 &3.3 & 3.9   $\pm$ 0.8  &    1.0 $\pm$ 0.2 &   2.5E4 &  1.E6 \\
  S29 &1.9 & 1.7   $\pm$ 0.3  &    0.4 $\pm$ 0.07 &   1.E3 &  2.E3 \\
  S30 &0.5 & 3.2   $\pm$ 1.0  &    0.8 $\pm$ 0.2 &   8.E3 &  3.E4 \\
  S34 &1.3 & 1.8   $\pm$ 0.4  &    0.4 $\pm$ 0.1 &   1.E3 &  2.E3 \\
             \noalign{\smallskip}
            \hline
           \end{tabular}
\\  The gas excitation is discussed in Sec \ref{excit}, 
 n(\hh) is given for T$_k$=60K and 30K, and N$_{\rm CO}/\Delta V$=7$\times$10$^{16}$ cm$^{-2}$/(\kms)
\end{table}

\subsection{Stellar mass and gas fraction}
\label{GFR}

The redshift evolution of the CO luminosity envelope found
in section \ref{CO-z} suggests that the declining gas content of
galaxies could be an important driver of the decline in the cosmic
star formation density since z=1.
To better quantify this, we now measure the gas fraction 
of all systems at different redshifts, and estimate their stellar masses.

\begin{figure}[h!]
\centering
\includegraphics[angle=-90,width=7cm]{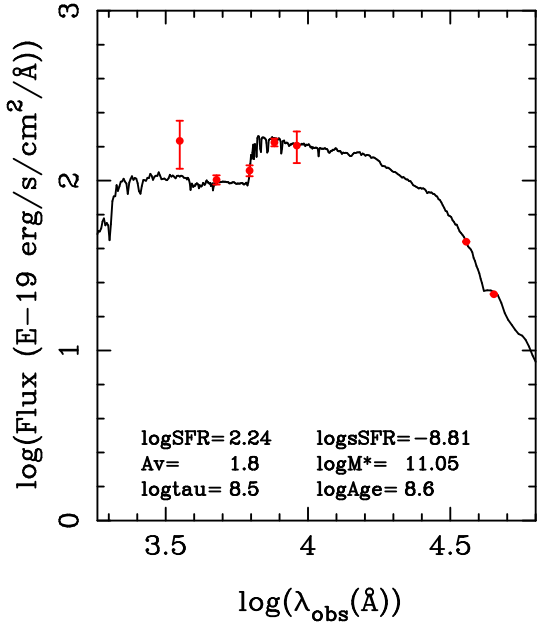}
\caption{Spectral energy distribution (SED) of S43
(shown here as a representative example), based on optical/near-IR broadband
photometry. The red points with error bars represent the observations
of SDSS (ugriz bands), and of Spitzer/IRAC 3.6 and 4.5\mics observations.
The black curve is the overplotted best-fit stellar synthesis model,
obtained with the FAST code (see text). An exponential star-formation history,
where time scale tau is assumed, with a Salpeter IMF. Values for 
tau and population age are given in yr, as well as stellar mass (\msol), visual extinction (mag) 
SFR (\msol/yr), or specific SFR (yr$^{-1}$). All values are in log, 
except A$_v$.}
\label{fig:plfit-43}
\end{figure}

To compute stellar masses from observed optical and near infrared (NIR) magnitudes, standard relations
exist as a function of colors, derived from stellar populations models
(see e.g. Bell \etal 2003).  We used them for the local z=0 galaxies. For higher
redshift objects, K-corrections need to be applied, and we used the analytical approximations from
Chilingarian \etal (2010) in Paper II. However, for z$>$0.5, these are no longer valid,
and for the present sample we estimated the stellar mass from SED-fitting 
of the optical and near-infrared luminosities taken from public catalogs, 
mainly the SDSS, 2MASS and Spitzer/IRAC fluxes. 
Most galaxies have only SDSS-DR8 fluxes, but ten have in addition
2MASS photometry, and another ten have Spitzer-IRAC fluxes.
The broadband photometry
were fitted using the code FAST (Fitting and Assessment of Synthetic Templates)
described in  Kriek \etal (2009). We selected the library of stellar
population synthesis models of Bruzual \& Charlot (2003), and adopted
an exponential SFR $\propto$ exp(-t/tau), with a
Salpeter initial mass function (IMF, Salpeter 1955). We used the
extinction law by Calzetti \etal (2000), and
metallicity was assumed to be solar.  As for the IRAC fluxes, only the 
first two channels (3.6 and 4.5 \mic) are relevant for the stellar component 
at the redshifts under consideration, so they were used 
for the fit.  The visual extinction A$_v$ was allowed to vary between 0 and 9, but
the best fits were obtained for values always lower than three. 
The model fitting also gives an estimation of the SFRs, which 
 frequently agrees within a factor 2-3 with what is derived from the 
far-infrared luminosities,
except in a few cases (where it can be an order of magnitude different). 
The time scale for star formation is
also derived, between the two limits that we set for the model fitting
 10 Myr $<$ tau $<$ 300 Myr. We imposed the latter upper limit, to be consistent with the starburst
nature of the objects. The parameters of the fit were selected such that the 
inverse of the specific SFR is always greater than tau.  
 An example fit for the source S43 is presented in Figure \ref{fig:plfit-43}.
  There is a certain degeneracy between the physical parameters, since the tau values
are not tightly constrained; however, the stellar mass is more robust, depending on the SED fit
and magnitudes observed. The derived masses are listed in Table  \ref{tab:lines}.

\begin{figure}[h!]
\centering
\includegraphics[angle=0,width=8cm]{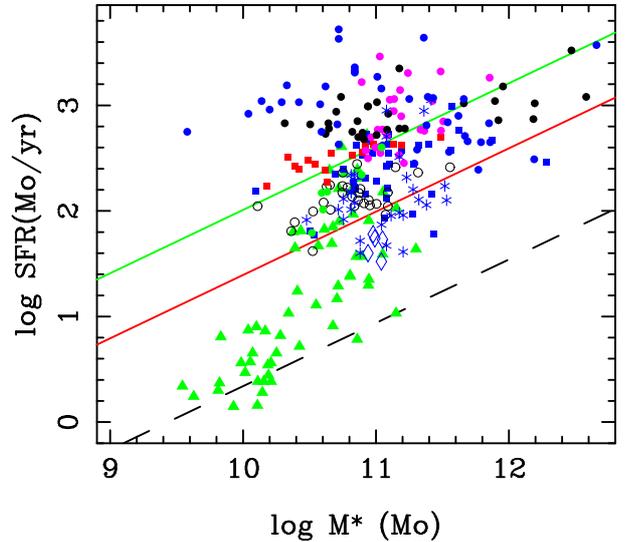}
\caption{The star formation rate (SFR) obtained from the far infrared luminosity,
versus stellar mass of galaxies in our sample (filled black circles for 0.2 $<$ z $<$ 0.6,
and filled blue circles for 0.6 $<$ z $<$ 1.0), compared
to  the sample of Da Cunha \etal (2010, filled red squares), Fiolet \etal (2009, filled magenta circles),
Gao \& Solomon (2004, filled green triangles), 
Solomon \etal (1997, filled blue squares), 
Chung \etal (2009, open circles),
Geach \etal (2009, 2011, blue diamonds),
Daddi \etal (2010, filled green circles), and
Genzel \etal (2010, blue asterisks).  The black dashed line
represents the star-forming galaxy main sequence at z=0, 
the red filled line the main sequence at z=1,
and the green filled line the main sequence at z=2.
%both have an adopted slope of 0.8 (Tacconi \etal 2010).}
All three lines have an adopted slope of 0.6 (Karim \etal (2011).}
\label{fig:SFR-Ms}
\end{figure}

Stellar masses lie between
4$\times$10$^{9}$ and 4$\times$10$^{12}$ M$_\odot$, with a median value of
 1.1$\times$10$^{11}$ M$_\odot$. 
Figure \ref{fig:SFR-Ms} displays the SFR, derived from the infrared luminosity,
versus stellar mass, in comparison with some other samples considered
before, and adding the ULIRG samples of Da Cunha et al (2010) 
and Fiolet et al (2009). 
It is interesting to locate the position on the graph of the main sequence
of star-forming galaxies, as defined by Noeske \etal (2007) and Daddi \etal (2007).
%We adopt the power laws of slope 0.8, and redshift evolution, as
%SFR$\propto$ M$_*^{0.8}$(1+z)$^{2.7}$ as used by Tacconi \etal (2010).
We adopt the power laws of slope 0.6, and redshift evolution, as
SFR$\propto$ M$_*^{0.6}$(1+z)$^{3.5}$ as found by Karim \etal (2011).
Our galaxy points  sample the region significantly above the main sequence,
with however some scatter, so that some galaxies reach the center
of the main sequence. The star-forming galaxies at z=0, i.e. green triangles in
Fig. \ref{fig:SFR-Ms} whose FIR luminosities
are not all ULIRG, are also somewhat above the z=0 main sequence. 

\begin{figure}[h!]
\centering
\includegraphics[angle=0,width=8cm]{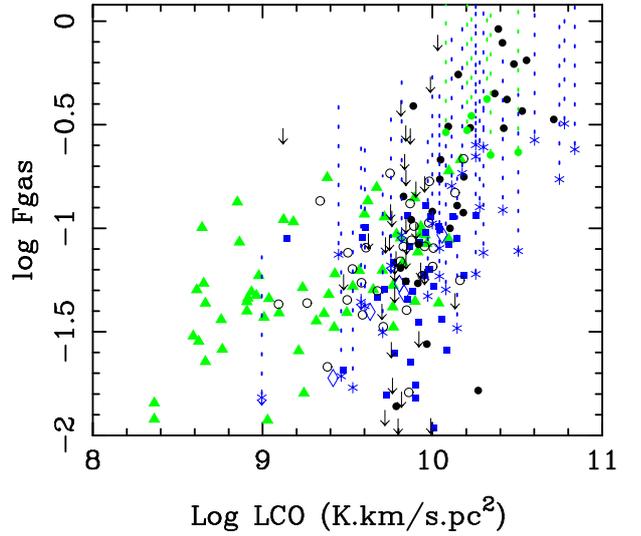}
\caption{Gas fraction=M{\hh}/(M(\hh)+M$_*$)
versus CO luminosity, assuming the same CO-to-\hh\, conversion factor
$\alpha$ = 0.8  M$_\odot$ (K \kms\, pc$^2$)$^{-1}$. 
The points of the high-z samples (blue asterisks and green dots),
have been continued by a dotted line joining the two extreme values of gas fraction,
obtained with $\alpha$=0.8 and 4.6.
 We indicate these lines only for the points in the high-z samples for 
which the MW conversion factor was selected by the authors.
All symbols are as defined in Fig \ref{fig:CO-z}. }
\label{fig:GFR-CO}
\end{figure}

\begin{figure}[h!]
\centering
\includegraphics[angle=0,width=8cm]{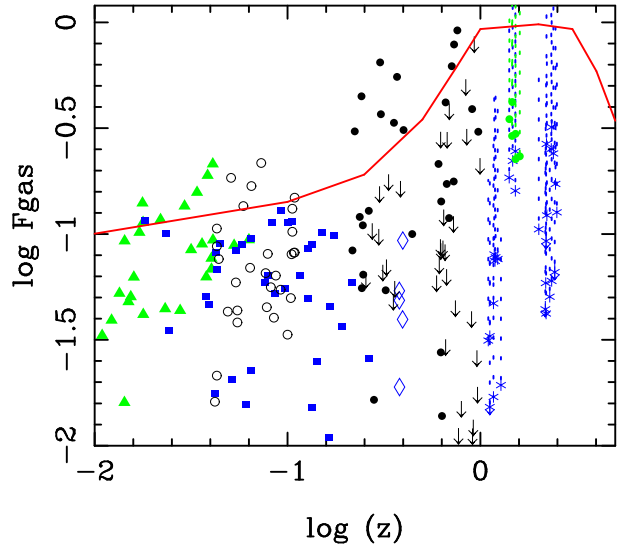}
\caption{The gas fraction as in Figure  \ref{fig:GFR-CO}, but versus redshift z.
The red curve is a schematic line summarizing the evolution of cosmic star formation density,
from the compilation by Hopkins \& Beacom (2006), complemented with the
GRB data by Kistler \etal (2009) and the optical data from Bouwens \etal  (2008).
The red curve is logarithmic and
only indicative of relative variations of the SFR per cubic Mpc
as a function of redshift, and can be translated vertically.
}
\label{fig:GFR-z}
\end{figure}

The gas fractions F$_{gas}$=M{\hh}/(M(\hh)+M$_*$) 
derived from our stellar masses 
show large variations (see Figure \ref{fig:GFR-CO}). 
  To obtain the gas fractions, we converted the CO luminosity to a gas mass using the
ULIRG factor ($\alpha$=0.8). For the high-z samples, i.e. blue asterisks and green dots
where the standard MW conversion has been selected by the authors, we extended in addition
the points by a dotted line joining the two extreme values of gas fraction,
obtained with $\alpha$=0.8 and 4.6 (the MW value).
  The gas fraction is well correlated with CO luminosity.
The average gas fraction in our galaxies is
 15\% for the 0.2$<$z$<$0.4 sample, and 24\% for the 0.6$<$z$<$1.0 sample.
For the low-z samples by Gao \& Solomon (2004), Solomon \etal (1997) and 
Chung \etal (2009), the averages are 5\%, 6\% and 7\%, respectively,
while they are 23\% and 10\% for the samples of Daddi \etal (2010) at z=1.5 and Genzel \etal (2010),
at z=1-2.
Assuming a standard MW conversion factor, the two last values become 63\% and 35\%.
  The increase in the gas fraction with z is clearly apparent,
as shown in Figure \ref{fig:GFR-z}.
To guide the eye, we have indicated the z-evolution of
the cosmic star formation history in this figure, as compiled by Hopkins \& Beacom (2006), from
different works in the literature, and complemented at very high redshift
by the gamma-ray burst (GRB) data of Kistler \etal  (2009)
and the optical data (LBG) from Bouwens \etal (2008).
In this approximate comparison, it
 is interesting to note that the rise in the SFR has only slightly higher amplitude
than the rise of gas fraction, supporting the hypothesis of the large role
of the gas in this evolution. 

\subsection{Activity of the galaxies}
\label{AGN-Int}

To determine what is triggering and regulating the starburst activity in our
sample galaxies, it is interesting to try to determine which galaxies are morphologically
perturbed or interacting, and to infer the activity status of their nucleus. 
 The last column of Table \ref{tab:h2} indicates the classification of
nuclear AGN activity and/or possible minor or major interactions.
 The latter have been derived from the available images, either from HST,
or presented by Stanford \etal (2000), or SDSS images. Some of the HST images of
galaxies detected in CO are shown in Figure \ref{fig:hst}, the rest in the Appendix.
  The strong or weak interactions were determined from the relative fraction
of the light in the tidal tails or perturbed features. No interaction means that 
the galaxy image looks unperturbed.

 At high infrared luminosities, the fraction of Seyfert is about 50\% in local ULIRGs, and their
fraction increases steeply with L$_{\rm FIR}$ (Veilleux \etal 1999). The tight link
between CO and infrared luminosities, however, shows that the dust heating is dominated
by the starburtst in these objects (Iono \etal 2009).
Among our 39 objects, 13 are known as AGN, and seven of these are detected in CO, implying
a detection rate of more than 50\%, higher than that of the whole sample.
 The percentage of weak and strong interactions in the whole sample are 23\% and 28\%, respectively,
while in the detected object sample they are both 33\%. There is thus a slight 
correlation between CO detection and interactions, however not at high significance.
  For more stringent conclusions, higher spatial resolution images need to be obtained for
the half of the sample without HST data.

\begin{figure*}[ht!]
\centering
\includegraphics[angle=-90,width=17cm]{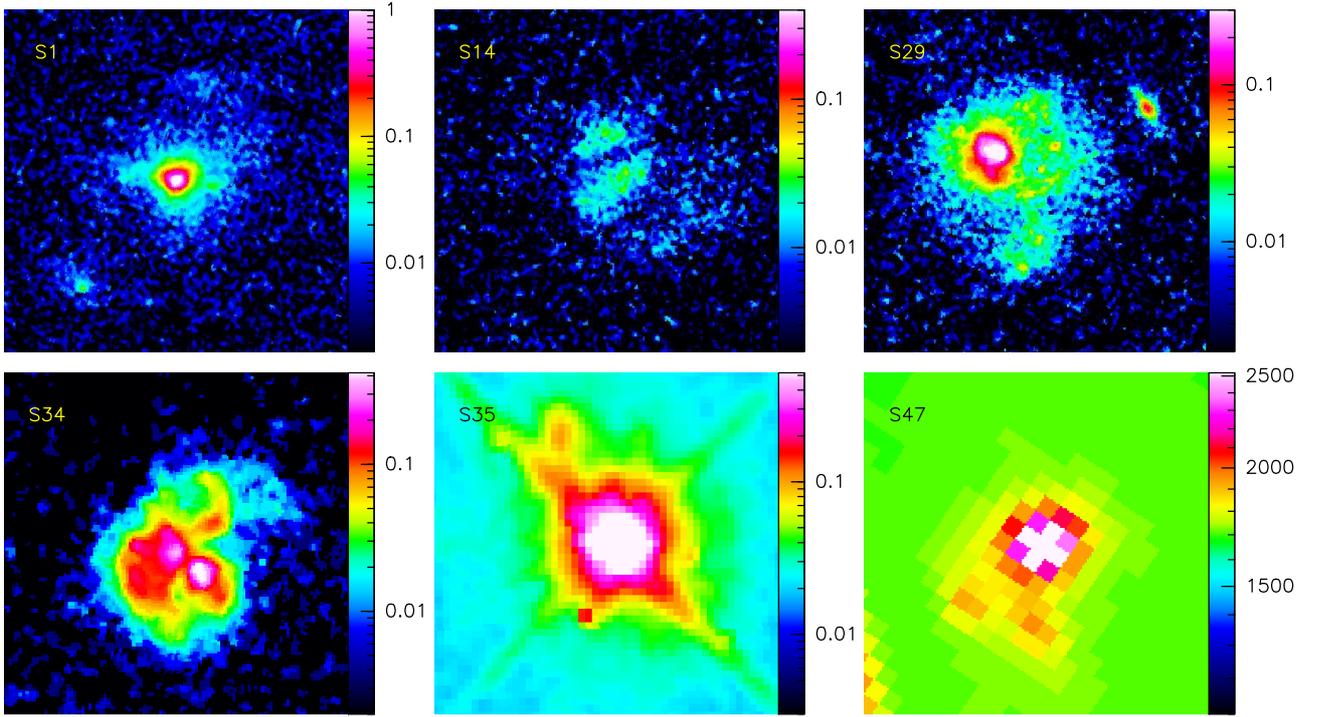}
\caption{Optical or NIR images from HST. Most images are ACS-F814W, except
S35 (WFPC2-702W) and S47 (WFC3-160W). 
All sources are detected in CO.
Each panel is 5$''$x5$''$ in size, and is centered
on the galaxy coordinates given in Table \ref{tab:sample}. The brightness scale
is logarithmic (except for S47).
North is up and east to the left in all panels.
S14, S29, S34 and S47 were classified as strongly interacting, while S1 and S35 are weakly interacting.
}
\label{fig:hst}
\end{figure*}

\subsection{Star formation efficiency}
\label{SFE}

\begin{figure}[h!]
\centering
\includegraphics[angle=0,width=8cm]{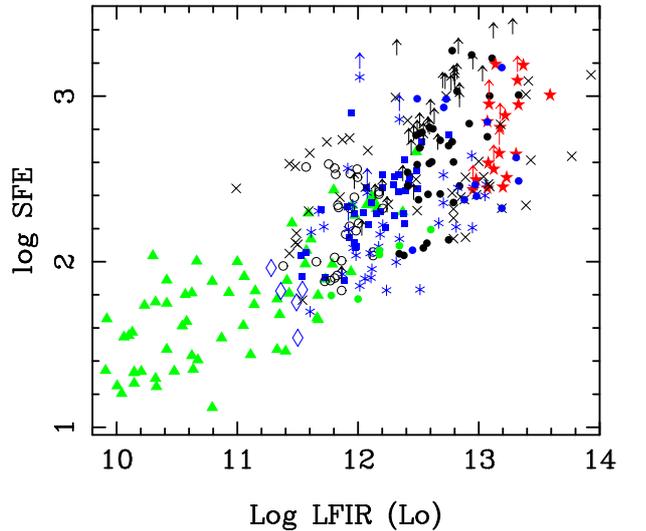}
\caption{Star formation efficiency SFE=L$_{\rm FIR}$/M(\hh), versus far-infrared
luminosity L$_{\rm FIR}$,  assuming a single CO-to-\hh\, conversion factor of
$\alpha$ = 0.8  M$_\odot$ (K \kms\, pc$^2$)$^{-1}$. All symbols are as defined
in Fig \ref{fig:CO-z}. }
\label{fig:SFE-FIR}
\end{figure}

\begin{figure}[h!]
\centering
\includegraphics[angle=0,width=8cm]{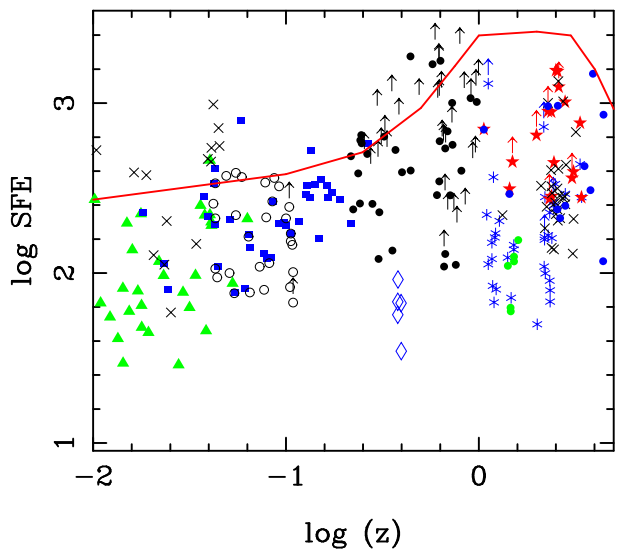}
\caption{Same as Figure  \ref{fig:SFE-FIR}, but versus redshift.  
The red line is the same as in Figure \ref{fig:GFR-z}.
The SFE shows comparable evolution with z as the gas fraction shows.
}
\label{fig:SFE-z}
\end{figure}

As in Paper II, we define the star formation
efficiency as SFE = L$_{\rm FIR}$/M(\hh),
assuming a constant CO-to-\hh\, conversion factor.
The average SFE in our 0.6 $<$ z $<$ 1 sample is 595 \lsol/\msol,
comparable to that of the 0.2$<$z$<$0.6 sample.
We plot SFE versus L$_{\rm FIR}$ in 
Figure \ref{fig:SFE-FIR}, and versus redshift in
Figure \ref{fig:SFE-z}.

Our intermediate-z sample shows some of the highest efficiencies
in star formation. The most extreme objects, with SFE $>$ 1000  \lsol/\msol\,
are mostly interacting or perturbed, such as S1, S2, S4, S5,
S11, S14, S20 and S25.
While there is a relatively good correlation between SFE and L$_{\rm FIR}$,
it is not the case for L'$_{\rm CO}$. The last shows only two populations
of objects, with the local starbursts at low efficiency and gas content, separated from
the higher-z samples. The SFE correlates much better with the dust temperature.
This was also found in Paper II, so we do not reproduce the figure here.

Finally, the evolution of SFE with redshift is shown
in Figure \ref{fig:SFE-z}. There is an obvious rise of the 
envelope between
z=0.2 and 1, precisely in the range of our sample.
 The amplitude of the rise appears, however, lower than 
that of the cosmic star formation
history, as compiled by Hopkins \& Beacom (2006). 
 Figure \ref{fig:GFR-z} shows that
the redshift variations of the gas fraction have a larger amplitude.  
  The star formation evolution is certainly due to a combination
of factors, essentially the gas fraction and the efficiency. Both 
vary with redshift in a similar manner, and it cannot be concluded 
which is the most determinant parameter

\begin{figure}[h!]
\centering
\includegraphics[angle=0,width=8cm]{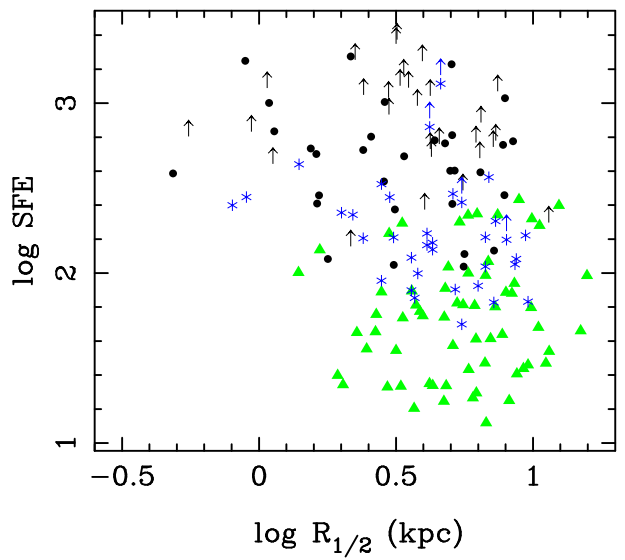}
\caption{Same as Figure  \ref{fig:SFE-FIR}, but versus half-light radius.
}
\label{fig:SFE-r}
\end{figure}

To search for a possible correlation between the SFE and the compactness
of the starburst, we computed the half-light radii of galaxies in our sample,
using the GALFIT software version 3.0.4 (Peng \etal 2002, 2010).  Red images in the I-band were
used at intermediate z, with the highest possible angular resolution: HST images when available,
and  SDSS images for the remaining galaxies. A single-component Sersic model was 
fit to all images, giving an
effective radius, which is displayed in Table \ref{tab:h2}. 
 With respect to compactness, it is interesting to compare our sample with
local star-forming galaxies. For that, we obtained  
the effective radii in the B-band for some local galaxies, from HyperLeda
(Paturel \etal 2003)\footnote{http://leda.univ-lyon1.fr}. This band is the best at z=0 to 
correspond with the I-band at intermediate redshift.
 The SFE is plotted versus 
the half-light radii of galaxies R$_{1/2}$, when available, in Figure \ref{fig:SFE-r}.
There is only a slight trend for SFE to be anticorrelated to R$_{1/2}$, or correlated to the
compactness of the blue light distribution, which reflects the star formation. 
 The compactness of the starburst might be better traced by the molecular gas distribution,
but this requires high resolution mapping of the high-z galaxies. 

\section{Discussion and conclusions}
\label{disc}

We have presented
 the completion of our survey of the gas content in galaxies at intermediate redshift,
through CO observations in a sample of 39 ULIRGs with redshifts between 0.6 and 1.0. 
Together with our earlier work (Paper II), this eliminates the CO desert in this
highly relevant epoch (0.2$<$z$<$1), in which the SFR in the universe
decreased by a factor 10. The CO detection rate between z=0.6 and 1 is 38\%,
significantly lower than  between z=0.2 and 0.6 (60\%, cf Paper II).
 The CO luminosity and, therefore, the derived \hh\, mass are increasing with
redshift by about a factor 4 up to z=1.  We estimated the stellar mass
of all sample objects through SED fitting of the optical and near-infrared
fluxes and derived gas fractions. The evolution of the gas fraction with redshift is
very pronounced, with a behavior reminiscent of the SFR evolution,
suggesting that the gas fraction plays a large role in determining the star formation
history.

\begin{figure}[h!]
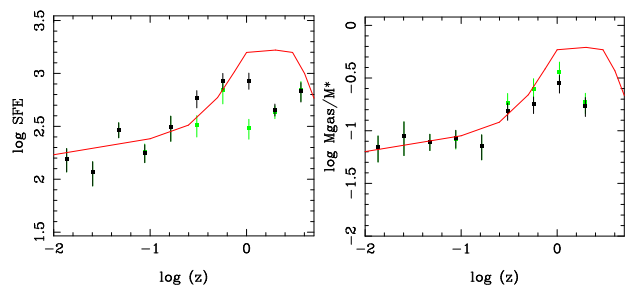

\centering
\includegraphics[angle=0,width=4cm]{SFE-bins.ps}
\includegraphics[angle=0,width=4cm]{gsratio-bins.ps}
\caption{Evolution with redshift of averaged quantities, SFE to the left, and gas-to-stellar mass ratio
to the right. The average of only detected points is plotted in green, and with the 3$\sigma$ upper limits in black
(for high-z samples only). The error bars are the statistical ones following the square root of the number of points averaged.
The red line is the same as in Figures  \ref{fig:GFR-z} and \ref{fig:SFE-z}.
}
\label{fig:SFE-bins}
\end{figure}

We also checked the evolution of star formation efficiency with redshift,
and  found  a trend comparable  to that of the gas fraction. 
We concluded that both parameters play a significant role in setting the SFH.
This can be seen in the envelope traced by  the most extreme objects, but also in 
the averages over all detected objects, with or without taking the upper limits into account
(Figure \ref{fig:SFE-bins}).
To quantify this important point better, we averaged all SFE values and gas-to-stellar mass ratios
over the redshift ranges z$<$0.2, and 0.2$<$ z $<$ 1.0, and computed the ratio between
 the two averaged values.  The latter depend on whether the averages include the 
upper limits. The SFE at intermediate redshifts experiences a jump by a factor  2.1 
(with detections only) and 3.8 (taking upper limits into account), while the 
 ratio of M$_{gas}$/M$_*$  rises by a factor  3.2 (with detections only) 
and 2.5 (taking upper limits into account). Given all these values,
we can conclude that both SFE and the gas-to-stellar mass ratio were
higher by a factor $3\pm1$ at 0.2$<$z$<$1.0.

We checked whether the SFE is related to the compactness of the starburst,
but found only a slight trend by SFE to decrease with half-light radius  (Figure \ref{fig:SFE-r}).

Since the gas fraction depends on the CO-to-H$_2$ conversion ratio adopted,
we observed a number of galaxies in both the CO(2-1) and CO(4-3) transitions to
derive the gas excitation. The latter varies from source to source, but we
find that in the majority the gas is significantly excited up to J=4, suggesting
a high \hh\, volume density and/or temperature. 
The comparison between the CO-derived gas mass and the dust mass
derived from the far infrared fluxes supports the choice of the ULIRG conversion 
factor for our sources. This does not exclude some of the sources
having low excitation, and their total gas mass has been underestimated.
Spatial information on the CO emission is needed to better constrain this issue,
and we will report the results of PdB
 interferometer imaging in a subsequent paper.

\begin{figure}[h!]
\centering
\includegraphics[angle=0,width=8cm]{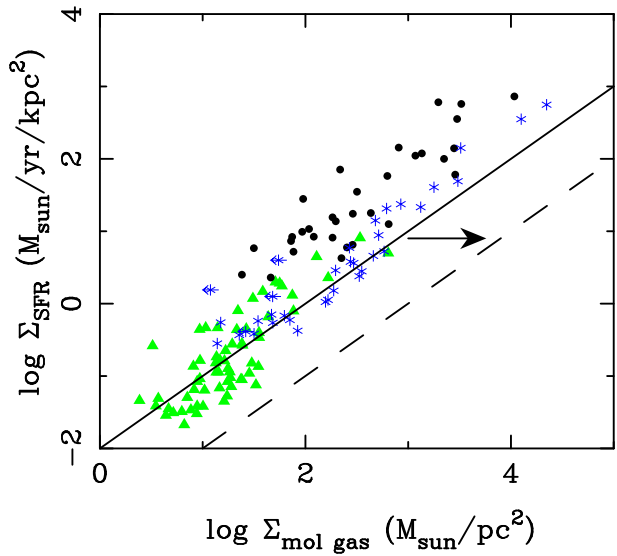}
\caption{ The Kennicutt-Schmidt relation between the gas surface
density and the SFR surface density for the galaxies of our sample
(filled black dots), compared to the galaxies from Gao \& Solomon (2004, 
filled green triangles), and from Genzel et al (2010, blue asterisks).
The full line corresponds to a depletion time of 100 Myr, and the dash line
to 1 Gyr. A common conversion factor of $\alpha$ = 0.8 has been adopted here.
The arrow indicates the direction and extent that all points will move,
were an MW conversion factor to be used.}
\label{fig:KS}
\end{figure}

One caveat could be the bias introduced by selecting ULIRGs
in our sample. It is true that our galaxies occupy the upper envelope
of the SFR diagrams, either as a function of stellar mass, as in Fig. 
 \ref{fig:SFR-Ms}, or as a function of molecular gas, as in 
Fig. \ref{fig:KS}. 
There is, however, a continuity in the various categories, and the depletion
time scales are progressive and overlapping, especially if the uncertainty
on the CO-to-\hh\, conversion factor is taken into account.
 This factor could vary smoothly across the observed galaxies, according
to the kinetic temperature of the gas, its volume density, and its
velocity dispersion. All these quantities could depend on the SFR, but also
on its distribution and compactness (e.g. Shetty \etal 2011, Narayanan
\& Hopkins 2012, Feldmann \etal 2012).

\begin{figure}[h!]
\centering
\includegraphics[angle=0,width=8cm]{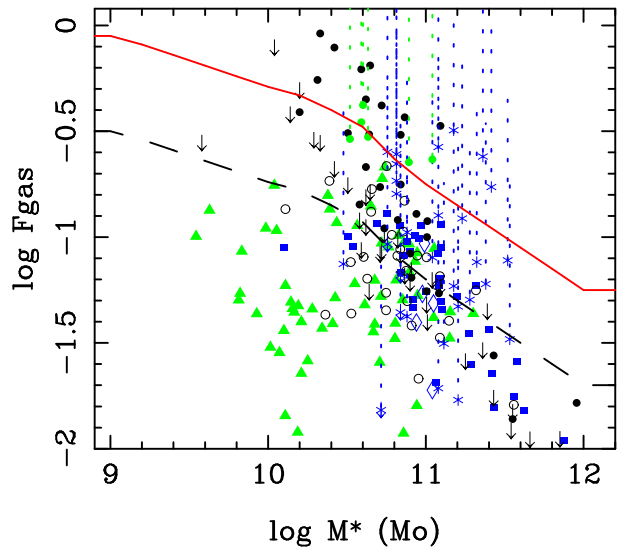}
\caption{The gas fraction as in Figure \ref{fig:GFR-CO}, but versus stellar mass.
The red line represents the predictions of cosmological simulations at z=1,
and the black dashed line at z=0 from Dav\'e \etal (2011).
}
\label{fig:GFR-Ms}
\end{figure}

Several physical parameters may intervene to account for the evolution
of galaxies since z=1. First the gas fraction is a key parameter, and we have shown that 
indeed, its evolution in this redshift range is significant. 
 The fueling of galaxies could be quenched by environmental effects,
such as gas stripping and strangulation of star formation, through group and cluster formation
during these epochs (e.g. Kimm \etal 2009). Also quenching could be morphological,
due to the growth of spheroids (Martig \etal 2009). In addition to supernovae, or
galactic winds, local photoionization of stars could regulate star formation
above a critical SFR, 
which depends on mass and redshift, such as to explain the decline of cosmic SFR 
between z=1 and 0 (Cantalupo 2010).
Second is the star formation efficiency,
which could be higher, because triggered either by galaxy interactions and mergers, which have been more
frequent in the past, but also through cold gas accretion, which is also thought to be efficient 
at these  epochs (Keres \etal 2005).
The last mechanism can also produce high gas fraction, resulting in 
instability-driven turbulence, perturbed disks, and clumpy gas distributions.
This could mimick galaxy interactions in the observable morphology.
Between z=1 and 0, hydrodynamical and semi-analytic studies predict
a quenching of star formation, beginning at high stellar mass, and  progressively
involving lower masses (e.g. Gabor \etal 2010, Dav\'e \etal 2011). 
It is interesting to compare observations
of the gas fraction versus mass, to better constrain the models, since they 
have not yet reached coherence with observations. Figure \ref{fig:GFR-Ms}
displays this relation, with the model lines superposed for z=0 and 1. 
They represent the best-fit models, corresponding to the no-wind simulation 
from Dav\'e \etal (2011).  Even if winds are neglected,
the gas fraction slowly decreases
with time because the intergalactic gas accretion rate decreases faster than 
the gas consumption rate in star formation.
The consideration of stellar winds and, in
particular, the momentum-driven winds are necessary to reproduce
metallicity evolution, 
 but all wind models lead to an under prediction of the gas fraction
in small galaxies. Other quenching mechanisms are also necessary for
massive galaxies, such as the influence of AGN feedback (Di Matteo \etal 2005). 

\begin{figure}[h!]
\centering
\includegraphics[angle=0,width=8cm]{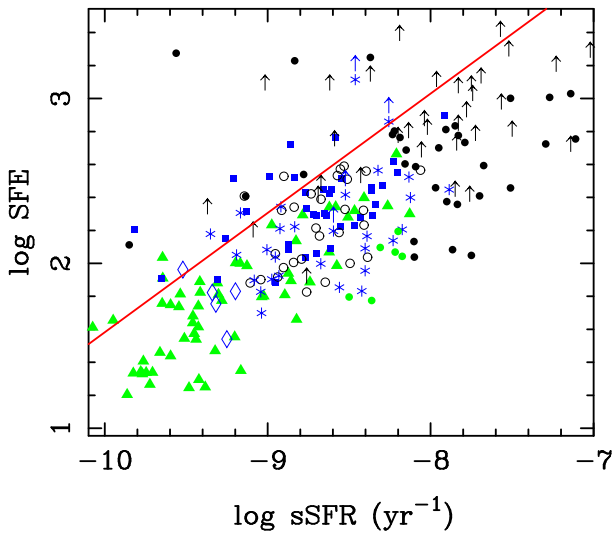}
\caption{ The star formation efficiency SFE (which varies as the inverse
of the depletion time scale for the molecular gas) versus the specific star 
formation rate sSFR, in yr$^{-1}$. The red line is the fit from the COLD GASS sample
of star-forming galaxies at z=0, from Saintonge \etal (2011b). 
}
\label{fig:SFE-sSFR}
\end{figure}

Another interesting parameter is the specific SFR,
or SFR normalized by the stellar mass,
which gives the time scale of formation of all the stellar mass in a galaxy
at the given present SFR.  We observe a good  correlation
with the star formation efficiency, in Figure \ref{fig:SFE-sSFR}.
We note, however, that the quantities on the vertical and horizontal axes
are not independent, since they both contain SFR.
This correlation has been studied, in particular, by Saintonge \etal (2011b),
who interpret the SFE as the  depletion time scale for the molecular gas.
 With respect to our definition of SFE, the depletion time scale t$_{dep}$ in yr is
such that log(t$_{dep}$) = 9.76 -log(SFE).
 The diagram helps to clearly identify the main sequence of star formation,
where galaxies have a continuous SFR, and the depletion time scale should
be comparable to the stellar-mass forming time scale.
 We have superposed the best fit line for the sample COLD GASS of 222
normal star-forming galaxies at z=0.  Our galaxies are generally below the line,
because of their larger gas content.

The observations discussed in this work cover most of the age of the universe (out to $z=1$) and 
virtually all of the increase in the cosmic SFR. 
 Both an increase in gas-to-stellar mass ratio and 
an increase in the SFE are responsible for the high cosmic SFR observed at earlier epochs.  
Depending 
on whether upper limits are taken into account, the M$_{gas}$/M$_*$ ratio
 increases by a factor 3.2 (with detections only) 
and 2.5 (taking upper limits into account) and the SFE rises by factors of 2.1 and 3.8, respectively.
 The two factors are therefore equally significant, and only the combination of the two
can explain the large increase in star formation between z=0 and 1.

%%%%%%%%%%%%%%%%%%%%%%%% acknowledgments
\begin{acknowledgements}
  We warmly thank the referee for constructive comments and suggestions. 
The IRAM staff is gratefully acknowledged for their
help in the data acquisition. FC thanks M. Kriek for providing her
IDL-based  FAST package for SED fitting, and acknowledges the
 European Research Council
for the Advanced Grant Program Num 267399-Momentum.
We made use of the NASA/IPAC Extragalactic Database (NED),
and of the HyperLeda database.
\end{acknowledgements}
%%%%%%%%%%%%%%%%%%%%%%%%%%%%%%%%%%%%%

\appendix
\section{Available images for more sources}

We present in this Appendix the remaining HST images of the sources,
which give insight into their morphology.

\begin{figure*}[ht]
\centering
\includegraphics[angle=-90,width=16.5cm]{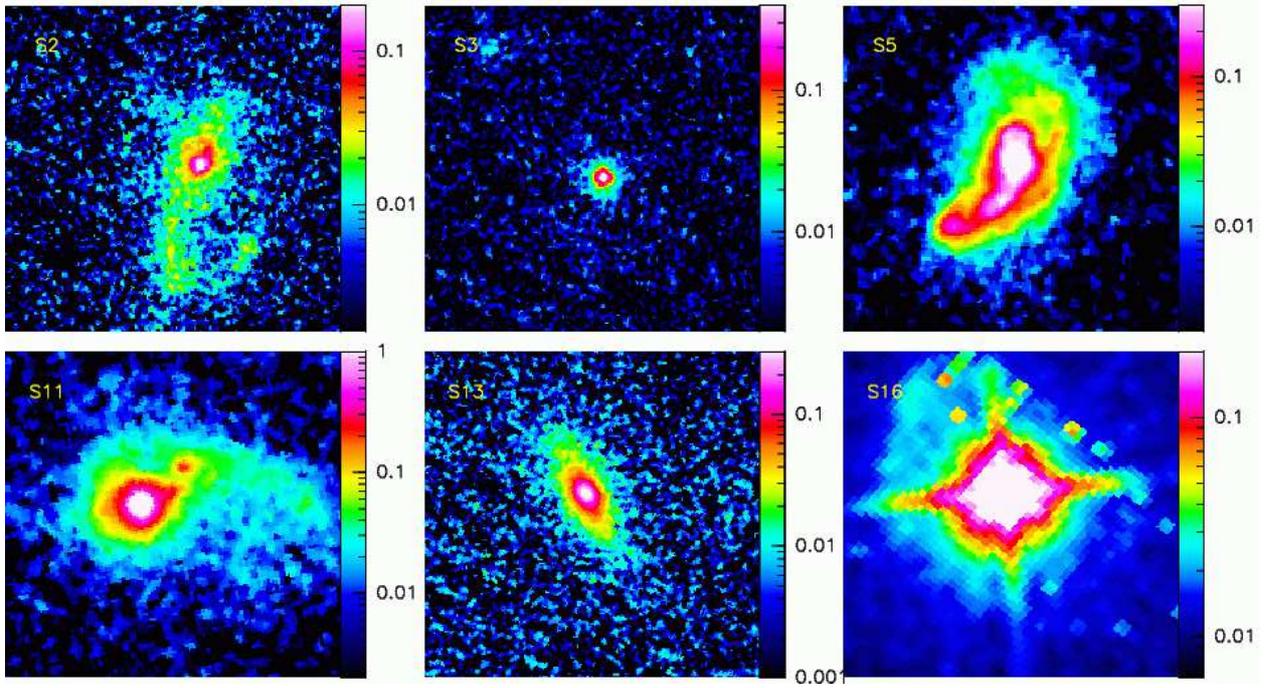}
\caption{Optical red  (F814W)  HST-ACS images of six more sources.
 Only S16 is detected in the CO line.
Each panel is 5$''$x5$''$ in size, and is centered
on the galaxy coordinates of Table \ref{tab:sample}. The brightness scale
is logarithmic.
North is up and east to the left in all panels.
}
\label{fig:hst1}
\end{figure*}

\begin{figure*}[ht]
\centering
\includegraphics[angle=-90,width=16.5cm]{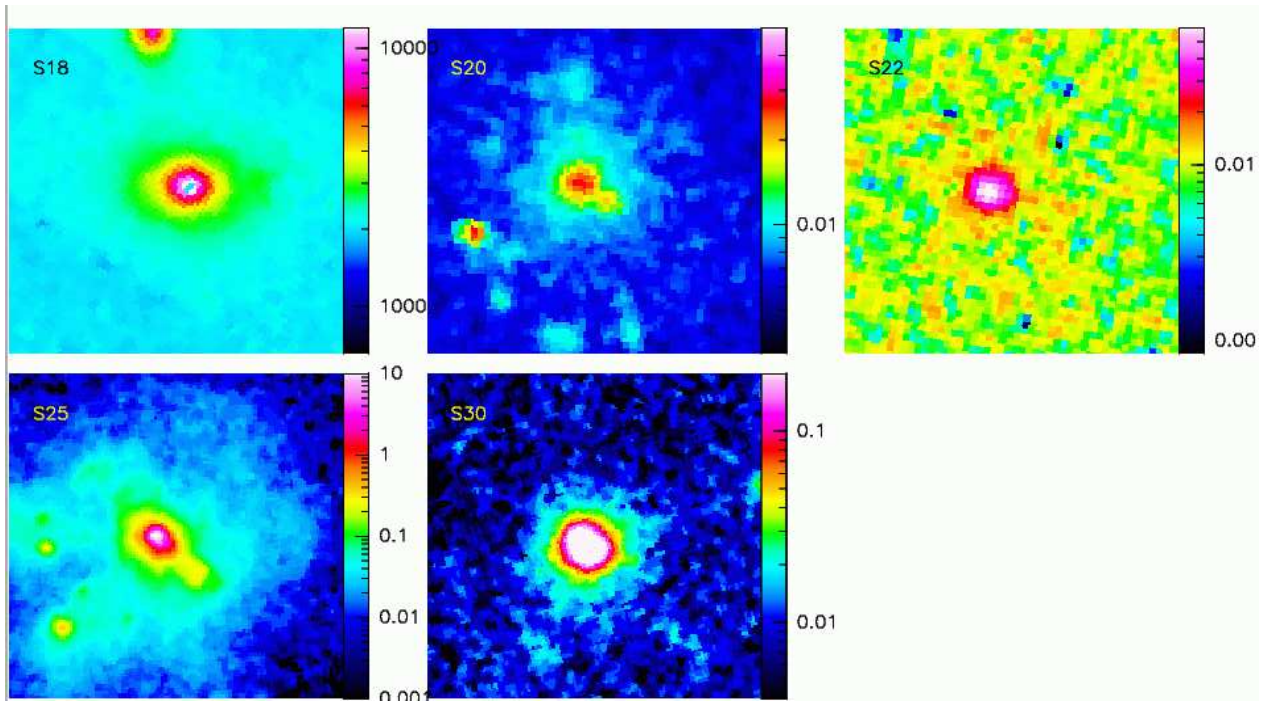}
\caption{ Same as in Fig \ref{fig:hst1}, except for  S18, which
is a NICMOS-F160W image. S22 and S30 are
detected in CO.
}
\label{fig:hst2}
\end{figure*}

\end{document}